\documentclass{article}
\usepackage[T1]{fontenc}
\usepackage[utf8]{inputenc}
\usepackage{geometry}
\usepackage{amsmath}
\usepackage{amsthm}
\usepackage{graphicx}
\usepackage{ae,lmodern}
\usepackage{caption}
\usepackage{subcaption}
\usepackage[authoryear]{natbib}
\usepackage{amsmath}
\usepackage{amssymb}
\usepackage{authblk}
\usepackage{tikz}
\usepackage{diagbox}
\usepackage{float}
\usepackage{caption}
\usepackage{verbatim}
\DeclareMathOperator{\tr}{tr}
\usepackage[colorinlistoftodos]{todonotes}
\usepackage[colorlinks=true, allcolors=blue]{hyperref}

\usepackage{lineno}
\usepackage{setspace}
\usepackage{soul}
\title{Effects of stage structure on coexistence: mixed benefits} 
\author[1,2,3,*]{Gaël Bardon}
\author[1,$\dagger$]{Frédéric Barraquand} 
\affil[1]{Institute of Mathematics of Bordeaux, University of Bordeaux, CNRS, Talence, France}

\affil[2]{Department of Polar Biology, Centre Scientifique de Monaco, Principality of Monaco}
\affil[3]{Department of Ecology, Physiology and Ethology, Institut Pluridisciplinaire Hubert Curien UMR 7178, University of Strasbourg, CNRS, Strasbourg, France}
\date{}

\begin{document}

\maketitle
\thispagestyle{empty}
%\linenumbers

\begin{abstract}
The properties of competition models where all individuals are identical are relatively well-understood; however, juveniles and adults can experience or generate competition differently. We study here less well-known structured competition models in discrete time that allow multiple life history parameters to depend on adult or juvenile population densities. A numerical study with Ricker density-dependence suggested that when competition coefficients acting on juvenile survival and fertility reflect opposite competitive hierarchies, stage structure could foster coexistence. We revisit and expand those results. First, through a Beverton-Holt two-species juvenile-adult model, we confirm that these findings do not depend on the specifics of density-dependence or life cycles, and obtain analytical expressions explaining how this coexistence emerging from stage structure can occur. Second, we show using a community-level sensitivity analysis that such emergent coexistence is robust to perturbations of parameter values. Finally, we ask whether these results extend from two to many species, using simulations. We show that they do not, as coexistence emerging from stage structure is only seen for very similar life-history parameters. Such emergent coexistence is therefore not likely to be a key mechanism of coexistence in very diverse ecosystems, although it may contribute to explaining coexistence of certain pairs of intensely competing species. 
\end{abstract}

~\\
\textbf{Keywords:} coexistence; stage structure; competition; matrix population models\\ 
~\\
\textbf{Correspondence:}
* \url{gael.bardon2@gmail.com}, 
$\dagger$ \url{frederic.barraquand@u-bordeaux.fr}\\

\newpage

\section{Introduction}
A basic tenet of demography and population ecology is that species vital rates, such as survival and fertility, can vary with age or stage \citep{leslie1945use, cushing1998introduction, caswell2001matrix}. Surprisingly, little of this rich age-structured population-level theory influences community-level coexistence theory, that by contrast mostly builds on unstructured Lotka-Volterra competition or consumer-resource  models \citep{chesson2000mechanisms,barabas2016effect,letten2017linking}. There have been, however, some calls to include more demography into community ecology \citep[e.g.,][]{miller2011thinking} and coexistence studies in particular, as well as a number of empirically-driven modelling studies doing so \citep{peron2012integrated,chu2015large}. They show in general that the vital rates of the various life-stages do not react in the same way to changes in the densities of juvenile and adult life-stages. These differences may well create new avenues for coexistence.  

Although mainstream coexistence theory often neglects the complexity of life histories, some stage-structured coexistence theory has been previously developed: several theoretical models, in past decades \citep{haigh1972can,loreau1994competitive} as well as in more recent literature \citep{moll2008competition, fujiwara2011coexistence} have considered complex life cycles in which ontogenetic changes occur, leading to models where vital rates can be differentially affected by competition. Such complexity can promote equalizing mechanisms \citep{fujiwara2011coexistence} as well as niche differentiation \citep{moll2008competition}. 
A main observation of \citet{moll2008competition} was the occurrence of so-called `emergent coexistence' in two-species two-stages models, which was defined as a coexistence equilibrium where competition on a single vital rate would lead to exclusion, but the combination of competition processes affecting two different vital rates leads to coexistence (e.g., competition affecting juvenile survival leads to species 1 winning, competition affecting fertility leads to species 2 winning, and yet both species coexist in the model with both types of competition). This prompted the exciting idea that community models with stage structure may foster more widespread coexistence than classic Lotka-Volterra theory, suggesting stage structure as a key missing ingredient in our explanations of diversity maintenance. 

However, the results of \citet{moll2008competition} had two limitations that prevented the generalization of this coexistence-enhancing role of stage structure. First, their investigations were restricted to a particular life cycle where juveniles and adults occupy separated habitats or niches (i.e., adults can only compete with adults and juveniles with juveniles), and these complex life cycles (sensu \citealt{wilbur1980complex}) were even suggested to be potentially responsible for the coexistence observed. Such segregation of competition faithfully represents some systems with metamorphosis, such as odonates (dragonflies and damselflies) or anurans (frogs), where there is an ontogenetic niche shift \citep{werner1984ontogenetic}. But such a model does not correspond to say, birds and mammals where adults can impose strong competition on the juveniles, as well as numerous other taxa of animals or plants for which the niche shift after development is only partial, which implies that the much larger adults can exert the bulk of the competition felt by juveniles. Thus whether competition acting on multiple life-history parameters could foster coexistence is still an open question for general life cycles, and whether complex life cycles have anything special that promotes coexistence is also something to uncover. Second, the analyses of \citet{moll2008competition} relied entirely on numerical simulations, which means that while we can observe emergent coexistence in the Ricker-based density-dependent models that they considered, it is still difficult to understand fully the phenomenon from a theoretical standpoint. Moreover, it is yet unclear if the emergent coexistence equilibria highlighted by \citet{moll2008competition} are structurally stable (robust to small deviations of the model framework, such as a slightly different shape of density-dependence) and realistic (occurring for a large range of parameter values). This highlights the need to investigate further structured population models with additional model structures and analytical techniques.

A further key question is whether emergent coexistence can occur in a  $S$-species context (where $S\gg2$). Indeed, the emergent coexistence observed by \citet{moll2008competition} resulted from a trade-off across the life-history stages which allowed each species to the best competitor for a given vital rate. While it is obvious that such a trade-off may occur with two species, there are multiple ways to generalize such trade-offs in competitive rankings to the $S$-species case, and not all of them necessarily lead to more coexistence. In fact, unstructured Lotka-Volterra models suggest great caution in generalizing results from two-species competition to many-species competition, as the criteria for coexistence with many species usually correspond to much broader niche separation between species than required for two species, with intraspecific competition dominating \citep{barabas2016effect}. Similar phenomena might be expected for stage-structured competition models.

Here, we extend the results of \citet{moll2008competition} to a new model framework using the canonical Beverton-Holt density-dependence (which is closer to the continuous-time logistic model, \citealp{cushing2004some}) and life cycles where juveniles can experience competition from adults. We present an analytical invasion criterion applicable in this new model as well as to the metamorphosis model structure numerically explored by \citet{moll2008competition}. This criterion allows us to better understand how and why emergent coexistence occurs in both models, and is valid whenever the single-species models have stable fixed points, which occurs for iteroparous life cycles. We also use the sensitivity analysis developed by \citet{barabas2014fixed} to explore the robustness of emergent coexistence equilibria. We find that emergent coexistence is a general feature of structured two-species competition models with two density-dependent vital rates, and that this equilibrium is not more sensitive to perturbation of parameters than `classic' coexistence equilibria allowed by niche separation. We finally explore $S$-species models' properties using simulations of diverse community structures (number of initial species, similarity of species) and for both life cycles. These many-species models show the limitation of the emergent coexistence mechanism to explain coexistence in highly speciose communities.

\section{Models and Methods}

\subsection{Competition models}

Our main model is a two-species difference equation with two life stages (juvenile and adult) for each species. It is a development of models presented in \citet{fujiwara2011coexistence}, which were restricted to the case where only one matrix parameter per species (either fertility or survival) is affected by densities. Our model extends this framework to multiple density-dependent parameters per species, which provides a useful springboard to combine two types of competition, affecting both fertility and survival rates. Our model could also be viewed as an extension to iteroparous life-histories and delayed maturation of the model of \citet{cushing2007coexistence}, who considered two-species competition with semelparous life-histories and fast maturation. Our model resembles \citet{moll2008competition}'s in that two vital rates are density-dependent, and in that the life cycle includes iteroparity and a delayed maturation; however, we consider like \citet{fujiwara2011coexistence} and \citet{cushing2007coexistence} Beverton-Holt functions to model competition. These competition models are sometimes called Leslie-Gower after \citet{leslie1958properties}, and are the closest discrete-time equivalents of continuous-time Lotka-Volterra competition models \citep{cushing2004some}. They promote fixed points equilibria, instead of cycles or chaos that can be generated by Ricker functions at high fecundities \citep{neubert2000density}, though see \citet{cushing2007coexistence} for semelparous life-histories and associated two-cycles. The Beverton-Holt functions help to find analytically the equilibria whenever multiple vital rates are density-dependent, which then fosters analytical insights with invasion criteria. 

In our main model (model 1), competition is generated by adults because it felt more realistic for many animal species in which adults and juveniles share the same habitat (in mammals and birds, see e.g. snowshoe hares populations, \citealp{boutin1984effect} or avian scavengers populations, \citealp{wallace1987competitive}). 

Without any competition, population densities of the two stages of a species after one time step can be expressed (we arbitrarily pick species 1 for illustrative purposes):
\begin{equation}
    n_{1j}(t+1) = n_{1j}(t) (1-\gamma_1)s_{1j} + n_{1a}(t) f_{1}
\end{equation}
and 
\begin{equation}
    n_{1a}(t+1) = n_{1j}(t) \gamma_1 s_{1j} + n_{1a}(t) s_{1a}
\end{equation}
where $\gamma_1$, $f_1$, $s_{1j}$ and $s_{1a}$ respectively denote maturation rate, fertility, juvenile survival rate, and adult survival rate of species 1, all in $[0,1]$ except $f_1 \in \mathbb{R}^{+}$.  Additionally requiring $\gamma_i>0$ and $s_{i,a}<1$ ($i \in \{1,2\}$) is necessary to ensure that individuals do not get infinite youth or infinite lifespans, respectively.

Then, we add competition on juvenile survival rates and fertility by making these parameters density-dependent. In our first and main model, we consider that competition is generated by adults of both species. Using the Beverton-Holt model of competition, we get:
\begin{equation}
    f_{1} = \frac{\pi_{1}}{1 + \alpha_{11} n_{1a} + \alpha_{12} n_{2a}}
\end{equation}
and

\begin{equation}
    s_{1j} = \frac{\phi_1}{1+\beta_{11}n_{1a}+\beta_{12}n_{2a}}
\end{equation}
where competition on fertility is composed of intra-specific competition ($\alpha_{11} n_{1a}$) and inter-specific competition ($\alpha_{12} n_{2a}$), and similarly for juvenile survival competition. Here we denoted $\phi_1$ and $\pi_1$ for maximal juvenile survival and fertility of species 1.

With the vector $\mathbf{n}_i = (n_{ij}, n_{ia})$, we have in matrix form $\mathbf{n}_i(t+1) = \mathbf{A}_i(\mathbf{n}(t))\mathbf{n}_i(t)$. Our model (model 1) can be written for one species with a projection matrix $\mathbf{A}_i$, for instance with $i=1$,
\begin{equation}
    \mathbf{A}_1(\mathbf{n}) = 
    \begin{pmatrix}
    \frac{(1-\gamma_1)\phi_1}{1+\beta_{11}n_{1a}+\beta_{12}n_{2a}} & \frac{\pi_{1}}{1 + \alpha_{11} n_{1a} + \alpha_{12} n_{2a} }\\
   \frac{\gamma_1\phi_1}{1+\beta_{11}n_{1a}+\beta_{12}n_{2a}}    & s_{1a}
    \end{pmatrix}
    =
    \begin{pmatrix}
    (1-\gamma_1)s_{1j}(\mathbf{n}) & f_1(\mathbf{n})\\
    \gamma_1 s_{1j}(\mathbf{n}) & s_{1a}
    \end{pmatrix}
\label{eq:combined_model}
\end{equation}
where $\alpha_{ij}$ and $\beta_{ij}$ are the competition coefficients associated to the effect of species $j$ on species $i$ on fertility and juvenile survival, respectively.

Although the mathematical derivations of equilibria do not impose that $s_{i,a}>0$ (iteroparity), we additionally impose that condition because the invasion and sensitivity analyses performed later require fixed point equilibria, while semelparous life-histories also produce more varied attractors, in particular 2-cycles \citep{cushing2007coexistence,cushing2012stable}.

We also considered a metamorphosis structure model (model 2) with a life cycle similar to the one described by \citet{moll2008competition} and \citet{cushing2007coexistence}, where juveniles compete only with juveniles and adults only with adults. Unlike \citet{moll2008competition} but similarly to \citet{cushing2007coexistence}, we chose Beverton-Holt density-dependence to model competition so that it can still be compared to model 1. However, our model 2 also differ from \citet{cushing2007coexistence}, who focus instead on a different life history context, with semelparity ($s_a = 0$), fast maturation ($\gamma=1$), and solely intraspecific competition on fertility ($\alpha_{21}=\alpha_{12} = 0$). In our model 2, juvenile survival does not depend on adult densities (as life-stages are separated) but depends on juvenile densities. 
We have a density-dependent juvenile survival rate of species $1$ given by:
\begin{equation}
    s_{1j}^{M}(t) = \frac{\phi_1}{1+\beta_{11}n_{1j}(t)+\beta_{12}n_{2j}(t)}.
\end{equation}
This model 2 can be written for one species with a projection matrix $\mathbf{A}_i^{M}$, for instance with $i=1$ and the same notations as for model 1:
\begin{equation}
    \mathbf{A}_1^{M}(\mathbf{n}) = \begin{pmatrix}
    \frac{(1-\gamma_1)\phi_1}{1+\beta_{11}n_{1j}+\beta_{12}n_{2j}} & \frac{\pi_1}{1+\alpha_{11}n_{1a}+\alpha_{12}n_{2a}} \\
    \frac{\gamma_1\phi_1}{1+\beta_{11}n_{1j}+\beta_{12}n_{2j}} & s_{1a}
    \end{pmatrix}.
\end{equation}\label{eq:MB_model}
We extended our models to $S$-species contexts, allowing each species to potentially interact with every other. For model 1, we obtained the following projection matrix (for species $i$):
\begin{equation}
     \mathbf{A}_i(\mathbf{n}) = 
    \begin{pmatrix}
    \frac{(1-\gamma_i)\phi_i}{1+\sum_{k=1}^{S}\beta_{ik}n_{k,a}} & \frac{\pi_{i}}{1 + \sum_{k=1}^{S}\alpha_{ik}n_{k,a}}\\
    \frac{\gamma_i\phi_i}{1+\sum_{k=1}^{S}\beta_{ik}n_{k,a}}    & s_{i,a}
    \end{pmatrix}.
    \label{eq:combined_model_Sspecies}
\end{equation}
The extension for model 2 is similar:

\begin{equation}
     \mathbf{A}_i^{M}(\mathbf{n}) = 
    \begin{pmatrix}
    \frac{(1-\gamma_i)\phi_i}{1+\sum_{k=1}^{S}\beta_{ik}n_{k,j}} & \frac{\pi_{i}}{1 + \sum_{k=1}^{S}\alpha_{ik}n_{k,a}}\\
    \frac{\gamma_i\phi_i}{1+\sum_{k=1}^{S}\beta_{ik}n_{k,j}}    & s_{i,a}
    \end{pmatrix}.\label{eq:combined_model_Sspecies_MB}
\end{equation}

\subsection{Scenarios of coexistence and associated parameter sets}\label{sec:param_sets}

We chose three scenarios of coexistence for illustration and sensitivity analysis, matching those described by \citet{moll2008competition}: (1) a \textit{classic} coexistence state that is indicated by both $\alpha$ and $\beta$ coefficients (equivalent to the usual intra $>$ interspecific competition coefficients in the factorised formulation of the unstructured Lotka-Volterra model, e.g. \citealt{letten2017linking}), (2) an \textit{emergent} coexistence state where the competition coefficients associated with the two vital rates indicate exclusion by the opposite species, and (3) a coexistence state where there is a priority effect indicated by competition coefficients acting on one vital rate while competition acting on the other vital rate indicates coexistence. For simplicity, we refer to these two last scenarios as `emergent coexistence' since coexistence is not guaranteed by the conditions verified separately by each set of coefficients. We note that coexistence scenario (3) does not strictly correspond to an emergent outcome as defined by \citet{moll2008competition} because competition coefficients for one of the two vital rates in fact already indicate coexistence.

Although we consider below arbitrary parameters in Invasion analysis, we also selected for numerical applications some parameter sets. To obtain parameter sets promoting emergent coexistence, we set a trade-off between $\alpha_{ij}$ and $\beta_{ij}$ for both species (large $\alpha_{ij}$ is associated to low $\beta_{ij}$ and vice versa).

To be able to compare the three scenarios, we set intraspecific competition coefficients to a constant and only changed interspecific competition coefficients. Note that we also chose the values of interspecific competition coefficients such that we obtain the same densities at equilibrium across the three coexistence scenarios. The equilibrium densities have been obtained through simulation of the dynamics over 3000 time steps. No closed-form solutions for the equilibrium densities are known for all possible values of parameters, but some analytical progress can be made in special cases (\ref{SI:fixedpoints_twospecies}). Parameter values for the three corresponding parameter sets are given in Table \ref{tab:full_3parameter_sets}. 

\begin{table}[H]
    \centering
\begin{tabular}{|c|rr|rr|rr|rr|rrrr|rrrr|}
%  \hline
Set & $\pi_1$ & $\pi_2$ & $\gamma_1$ & $\gamma_2$ & $\phi_1$ & $\phi_2$ & $s_{1a}$ & $s_{2a}$ & $\alpha_{12}$ & $\alpha_{21}$ & $\beta_{12}$ & $\beta_{21}$ \\ 
  \hline
1 & 30 & 25 & 0.7 & 0.8 & 0.5 & 0.4 & 0.5 & 0.6 & 0.05 & 0.06 & 0.06 & 0.06\\ 
2 & 30 & 25 & 0.7 & 0.8 & 0.5 & 0.4 & 0.5 & 0.6 & 0.02 & 0.112 & 0.125  & 0.01\\ 
3 & 30 & 25 & 0.7 & 0.8 & 0.5 & 0.4 & 0.5 & 0.6 & 0.043 & 0.035 & 0.155  & 0.165\\ 
%   \hline
\end{tabular}
\caption{Parameter sets chosen to obtain equal equilibrium densities but corresponding to contrasted scenarios of coexistence. We use intraspecific coefficients $\alpha_{ii}=\beta_{ii}=0.1$.}
\label{tab:full_3parameter_sets}
\end{table}

The three parameter sets led to the same densities at equilibrium with $n_1 = 200\pm1$ and $n_2 = 130\pm1$ (these have been fine-tuned by changing interspecific interactions while keeping other parameters constant). 

To check what coexistence outcomes were indicated by the competition coefficients associated to fertility and juvenile survival, we used the invasion criteria given by \citet{fujiwara2011coexistence}, valid for models with a single density-dependent vital rate affected by competition (either fertility or juvenile survival). For example, the invasion criteria of the model with only fertility affected by competition (i.e., with only $\alpha$ coefficients) is given by:
\begin{equation}
\mathcal{R}_\alpha(i) = \frac{R'_i}{R'_j} \frac{\alpha_{jj}}{\alpha_{ij}} 
\end{equation}
where 
\begin{equation}
    R'_i = \frac{\pi_{i}}{f_i^{(c)}} - 1\\
\end{equation}
and
\begin{equation}
    f^{(c)}_i = \frac{1}{\gamma_i s_j}(1 - s_a) \left(1- s_j+ \gamma_i s_j\right).
\end{equation}
This criterion $\mathcal{R}_{\alpha}$ reflects invasion outcomes if we only had density-dependence on fertility,  $\mathcal{R}_{\beta}$ does the same if we only had density-dependence on juvenile survival.
These invasion criteria aim to confirm rigorously if, based on $\alpha$ competition coefficients only, or $\beta$ coefficients only, we should conclude to coexistence or exclusion (the well-known rule intrasp. competition $>$ intersp. competition only works in certain parameterisations of unstructured two-species Lotka-Volterra models, not structured ones). As shown in Table \ref{tab:3param_sets_inv_crit}, the three parameter sets do correspond to the three mechanisms mentioned above. Parameter set 1 corresponds to mutual invasibility in models with a single vital rate that is density-dependent. Parameter set 2 matches emergent coexistence since 1 excludes 2 when competition is on fertility ($\alpha$ coefficients) while 2 excludes 1 when competition is solely on juvenile survival ($\beta$ coefficients). With parameter set 3, we have classical coexistence through mutual invasibility indicated by the model with competition on fertility only ($\alpha$ coefficients) and a priority effect in the model with competition on juvenile survival only ($\beta$ coefficients). 
\begin{table}[h]
\centering
\begin{tabular}{|c|c|c|c|c|c|}
  %\hline
Parameter set & $\mathcal{R}_\alpha$ & $\mathcal{R}_\beta$ & Outcome \\ 
  \hline
Set 1 & \textbf{2.29} / \textbf{1.46} & \textbf{1.76} / \textbf{1.58} & Classical coexistence \\ 
Set 2 & \textbf{5.72} / 0.78 & 0.84 / \textbf{9.47} & Emergent coexistence \\ 
Set 3 & \textbf{2.60} / \textbf{2.43} & 0.68 / 0.57 & Emergent coexistence \\ 
   %\hline
\end{tabular}
\caption{Invasion criteria for species 1/species 2, evaluated for models where a single vital rate is affected by competition. $\mathcal{R}_\alpha$ and $\mathcal{R}_\beta$ are invasion criteria respectively for models with competition only on fertility and juvenile survival. Invasion criteria whose values are larger than 1 are in bold.}
\label{tab:3param_sets_inv_crit}
\end{table}

\subsection{Invasion analysis of 
exclusion equilibria}\label{sec:mat_met_inv_analysis}

We devised analytical invasion criteria for our model combining competition on fertility and juvenile survival, starting with model 1 where all competition is generated by adults. These differ from the invasion criteria of the previous section in that both $\alpha$ and $\beta$ are now potentially non-zero. We were inspired by the work of \citet{cushing1998introduction, cushing2008matrix} who studied the stability of the exclusion equilibria (with the focal invading species absent) to describe the ability of the focal species to invade a community. We therefore looked for conditions equivalent to an unstable exclusion equilibrium, which corresponds to the fact that the absent species can actually invade. All equations are presented in \ref{app:inv_analysis}.

We used the aggregated parameters $C= (1-\gamma)\phi$ and $D= \frac{\pi\gamma\phi}{1-s_a}$ in the single-species model. By considering the model with a single species (since at the time of invasion, the invader is almost absent), we found an expression for the inherent net reproductive number, $C+D = (1-\gamma)\phi +\frac{\pi \gamma \phi}{1-s_a}$. This number must exceed 1 to have a viable population of the resident species. We then obtained the densities at a stable positive equilibrium for the single-species model:
\begin{equation}
    \left\{
\begin{array}{r c l}
n_j^* &=&  (1-s_a)\left(\frac{1+\beta n_a^*}{\gamma \phi}\right)n_a^*\\
n_a^* &=& \frac{(\alpha C-\alpha-\beta) + \sqrt{(-\alpha C + \alpha + \beta)^2 - 4\alpha\beta(-C-D+1)}}{2\alpha\beta}
\end{array}
\right.
\end{equation}
provided that
\begin{equation}
   C+D = (1-\gamma)\phi +\frac{\pi \gamma \phi}{1-s_a} > 1. 
\end{equation}
We then evaluated the stability of the exclusion equilibrium in the two-species model: if it is locally asymptotically stable, the invader converges to zero density; if not, the invader invades. To do this, we calculated the eigenvalues of the Jacobian of the 2 species system evaluated at the exclusion equilibrium. We found the following condition for stability when species 2 is absent and species 1 is resident:
\begin{equation}\label{eq:invasion_criteria}
    \frac{C_2}{1+\beta_{21}n_{1a}^*}+\frac{D_2}{(1+\beta_{21}n_{1a}^*)(1+\alpha_{21}n_{1a}^*)} < 1
\end{equation}
with $C_2= (1-\gamma_2)\phi_2$ and $D_2= \frac{\pi_2\gamma_2\phi_2}{1-s_{2a}}$ as denoted previously. This last expression (eq.~\ref{eq:invasion_criteria}) gives us an invasion criterion, in the sense that if the expression is larger than 1, species 2 is able to invade the community when rare. 

We applied the exact same method to model 2 with metamorphosis structure, where juveniles compete with juveniles and adults compete with adults (\ref{SI:inv_analysis_MB}).

We found analytical expressions for the boundaries of coexistence domains over $\alpha_{12}$ and $\alpha_{21}$. These expressions are respectively given by equations \ref{eq:2spcoExcondition1} and \ref{eq:2spcoExcondition2} that are derived directly from the invasion criteria:
\begin{align}
    \alpha_{21}^* &= \left( \frac{D_2}{(1+\beta_{21}n_{1a}^*)-C_2}-1\right)\frac{1}{n_{1a}^*}\label{eq:2spcoExcondition1}\\
    \alpha_{12}^* &= \left( \frac{D_1}{(1+\beta_{12}n_{2a}^*)-C_1}-1\right)\frac{1}{n_{2a}^*}\label{eq:2spcoExcondition2}
\end{align}
If both conditions $\alpha_{12}$<$\alpha_{12}^*$ and $\alpha_{21}$<$\alpha_{21}^*$ are met, then there is coexistence. If both conditions are violated, there is a priority effect, and if one condition is met and the other is violated, there is exclusion of one of the species.

To visualize a case of emergent coexistence outcomes as did \citet{moll2008competition} for their Ricker competition model, we represent the coexistence domains over $\alpha_{12}$ and $\alpha_{21}$ with fixed values of $\beta_{12}$ and $\beta_{21}$, for model 1. We stress that here, these coexistence domains arise from analytical expressions of the boundaries rather than numerical simulations of the dynamical systems. We computed $\alpha_{12}^*$ and $\alpha_{21}^*$ for the three parameter sets mentioned in section \ref{sec:param_sets} where $\beta$ coefficients indicated respectively coexistence, exclusion of species 1 and a priority effect. 

\subsection{Sensitivity analysis}

To examine the sensitivity of equilibrium densities to changes in parameters under different scenarios of coexistence, we applied the community-level sensitivity analysis developed by \citet{barabas2014fixed} to model 1. The computation of fixed point densities, required to perform the analysis, was done numerically.  With iteroparity, only stable fixed points were observed, both in single-species and two-species competition models, as shown in \ref{SI:simuls_twospecies}. The two-species competition models fixed points do not have closed-form expressions in the general case---although this may sound surprising (but see \citealt{meszena1997adaptive}), it was indeed easier to compute invasion criteria than to find analytically the two-species fixed point, that may be analytically tractable only in special cases (see \ref{SI:fixedpoints_twospecies}). We sum up here the notations and expressions used to compute the sensitivity. Explaining all the steps would require paraphrasing too much of \citet{barabas2014fixed}, so we simply refer the reader to their paper for more details.
We use $n_i = \mathbf{n}_i\mathbf{q}_i$ where $\mathbf{n}_i$ is the vector of population density of the species $i$ at equilibrium for the various stages and $\mathbf{q}_i$ a vector of weights given to stages. We considered only the case where $\mathbf{q}_i$ is a vector of ones. We also need to define $\mathbf{p}_i = \frac{\mathbf{n}_i}{n_i}$ the proportion of each stage of species $i$.
We denote $\mathbf{A}_i$ the projection matrix of species $i$ with its eigenvalues $\lambda^k_i$ and its left and right eigenvectors $\mathbf{v}_i^k$ and $\mathbf{w}_i^k$. We also define the dominant eigenvalue associated with its eigenvector : $\lambda_i$ and $\mathbf{w}_i$. We normalized the vectors to have $|\mathbf{w}_i|=1$ and $\mathbf{v}_i^j \mathbf{w}_i^k = \delta_{jk}$ with $\delta_{jk} = 1$ if $j=k$ and $0$ otherwise.
\citet{barabas2014fixed} used regulating variables $\mathcal{R}_\mu$ which correspond for example for species 1 in our model 1 to $\mathcal{R}_{\alpha,1} = \alpha_{11}n_{1a} + \alpha_{12}n_{2a}$ and $\mathcal{R}_{\beta,1} = \beta_{11}n_{1a} + \beta_{12}n_{2a}$. Note that the notation for regulating variables of \citet{barabas2014fixed} is similar to that used for invasion criteria in single-species models that we previously introduced, even though ecological meanings differ greatly. As regulating variables are not used later in the text, there is no risk of confusion and we simply follow the conventions of \citet{barabas2014fixed} in this section.

We then computed the sensitivity of population densities of both species to perturbations of each parameter using the following expressions, where $E$ is the perturbed parameter:
\begin{equation}\label{eq:sensitivity_expression}
    \frac{dn_i}{dE} = \sum^S_{j=1}(M_{ij})^{-1}g_j
\end{equation}

where the matrix $M_{ij}$ is given by
\begin{equation}\label{eq:matrix_mij}
    M_{ij} = \sum_{\mu, \nu}\left(\mathbf{v}_i\frac{\partial\mathbf{A}_i}{\partial\mathcal{R}_\mu}\mathbf{w}_i\right)\left(\delta_{\mu\nu}-\frac{\partial\mathcal{G}_\mu}{\partial\mathcal{R}_\nu}\right)^{-1}\frac{\partial\mathcal{R}_\nu}{\partial\mathbf{n}_j}\mathbf{p}_j
\end{equation}

and the vector $g_j$ is given by 
\begin{equation}\label{eq:vector_gj}
    g_j=\mathbf{v}_j\frac{\partial\mathbf{A}_j}{\partial E}\mathbf{w}_j + \sum_{\sigma, \rho}\left(\mathbf{v}_j\frac{\partial\mathbf{A}_j}{\partial\mathcal{R}_\sigma}\mathbf{w}_j\right)\left(\delta_{\sigma\rho}-\frac{\partial\mathcal{G}_\sigma}{\partial\mathcal{R}_\rho}\right)^{-1}\frac{\partial\mathcal{G}_\rho}{\partial E}.
\end{equation}
The function $\mathcal{G}_\mu(\mathcal{R}_\nu,E)$, introduced by \citet{barabas2014fixed} to simplify notations, is given by
\begin{equation}\label{eq:G_mu}
    \mathcal{G}_\mu(\mathcal{R}_\nu,E) = \sum_{j=1}^{S}\left(\frac{n_j}{\mathbf{q}_j\mathbf{w}_j}\frac{\partial\mathcal{R}_\mu}{\partial\mathbf{n}_j}\sum^{s_j}_{k=2}\frac{1}{\lambda_j-\lambda_j^k} \times \left(\mathbf{w}^k_j-\frac{\mathbf{q}_j\mathbf{w}^k_j}{\mathbf{q}_j\mathbf{w}_j}\mathbf{w_j}\right)\otimes\mathbf{v}_j^k\right)\mathbf{A}_j(\mathcal{R}_\nu,E)\mathbf{w}_j.
\end{equation}

With these expressions, we were able to compute the sensitivity to perturbations of parameter values for the three parameter sets / contrasted scenarios of coexistence. These have near-equal densities at equilibrium (see section \ref{sec:param_sets}), to allow for meaningful comparisons.

\subsection{$S$-species simulations}

We simulated the dynamics of larger communities (5, 10 and 40 species) with parameters chosen to favour emergence of coexistence through life-history complexity, using the extension of model 1 and of model 2 in $S$-species contexts (eqs. \ref{eq:combined_model_Sspecies} and \ref{eq:combined_model_Sspecies_MB}). Our idea to extend the mechanism behind emergent coexistence from two to an arbitrary number of species was to create opposite competitive hierarchies in the competition coefficients related to fertility ($\alpha$) and juvenile survival ($\beta$). We therefore devised a mechanism by which a species strongly competitive on fertility would be weakly competitive on juvenile survival and reciprocally. This was done through a negative correlation between the coefficients $\alpha_{ij}$ and $\beta_{ij}$. The negative correlation was parameterized using a bivariate normal distribution and a covariance between the two variables equal to $-0.9\sigma^2$, where $\sigma^2$ is the variance of both $\alpha_{ij}$ and $\beta_{ij}$. 

To draw the full set of parameter values, we have used:
\begin{itemize}
    \item Log-normal fertilities $\pi_i \sim \text{Log-}\mathcal{N}(\mu_\pi,\sigma_\pi^2)$
    \item Beta-distributed survival and transition probabilities $\phi_i \sim \mathcal{\text{Beta}}(a_\phi, b_\phi)$, $s_{a,i} \sim \mathcal{\text{Beta}}(a_{s_{a}}, b_{s_{a}})$ and $\gamma_i \sim \mathcal{\text{Beta}}(a_\gamma, b_\gamma)$
    \item Normally distributed intraspecific competition coefficients $\alpha_{ii} \sim \mathcal{N}(\mu_{\alpha},\sigma_{\alpha}^2)$ and $\beta_{ii} \sim \mathcal{N}(\mu_{\beta},\sigma_{\beta}^2)$
    \item As mentioned above, when $i \neq j$ we use interspecific coefficients drawn as $(\alpha_{ij},\beta_{ij}) \sim \mathcal{N}_2(\mu_{\alpha\beta},\Sigma_{\alpha\beta})$
\end{itemize}

with $\Sigma_{\alpha\beta} = 
\begin{pmatrix}
\sigma^2 & \rho \sigma^2 \\ \rho  \sigma^2 & \sigma^2
\end{pmatrix} $ and $\rho = -0.9$. We have $\mu_{\alpha\beta} = 0.8 \begin{pmatrix} \mu_{\alpha}\\ \mu_{\beta} \end{pmatrix}$ to ensure a strong interspecific competition, which seemed necessary to see possible effects of trade-off between competitive coefficients. Meta-parameter values determining the abovementioned parameter distributions are given in full in \ref{app:S_species_simus}. We chose 3 sets for the distribution-level parameters ($\mu_\pi,\sigma_\pi,a_\phi, b_\phi,...$),  corresponding to 3 different levels of variance in the vital rates that allowed us to have either species with very close vital rates (i.e., low variance of the parameter distributions across species) or more variable vital rates (medium and high variation, see \ref{app:S_species_simus}).

We simulated the dynamics and computed the number of extant species after 3000 time steps, which allowed to reach an equilibrium state (see  \ref{app:S_species_simus}). Then, for each of the three main parameter sets, we created new, replicated parameter sets where we permuted all the inter-specific coefficients in order to remove the correlation between $\alpha_{ij}$ and $\beta_{ij}$, while keeping the same inter-specific competition values. We performed these permutations 200 times and represented the histogram of the number of species that coexist after 3000 time steps for the 200 permutated datasets. This allowed us to compare scenarios where $\alpha_{ij}$ and $\beta_{ij}$ are negatively correlated to the null hypothesis of no correlation between them (which renders the emergence of coexistence through stage structure impossible). 

\section{Results}

\subsection{Invasion analysis}

First of all, the analytical expressions for the boundaries of coexistence domains obtained thanks to the invasion analysis (eq.~\ref{eq:2spcoExcondition1} and \ref{eq:2spcoExcondition2}) can be interpreted to understand how the trade-off between competition coefficients actually promotes coexistence. In equation \ref{eq:2spcoExcondition1} (and similarly in equation \ref{eq:2spcoExcondition2}), as $n_{1a},C_2,D_2$ are independent of inter-specific competition coefficients, a decrease of $\beta_{21}$ (the effect of species 1 on species 2's juvenile survival) on the right hand side will mechanically increase the value of $\alpha^*_{21}$ on the left-hand site (the boundary value for $\alpha_{21}$, the effect of species 1 on species 2's fertility). What does this mean ecologically? When the effect of species 1 on species 2's juvenile survival $\beta_{21}$ decreases, $\alpha_{21}$ can increase without species 2 getting extinct. This means that as one type of competition is lowered, the other type of competition can be increased while having both species coexisting. Thus we can have coexistence with low $\beta_{21}$ and high $\alpha_{21}$, and reciprocally, using the other equation, with high $\beta_{12}$ and low $\alpha_{12}$. That is, we have coexistence through opposite competitive hierarchies, which can be obtained for the full range of competition coefficients.   

\begin{figure}[H]
    \centering
    \includegraphics[scale=0.8]{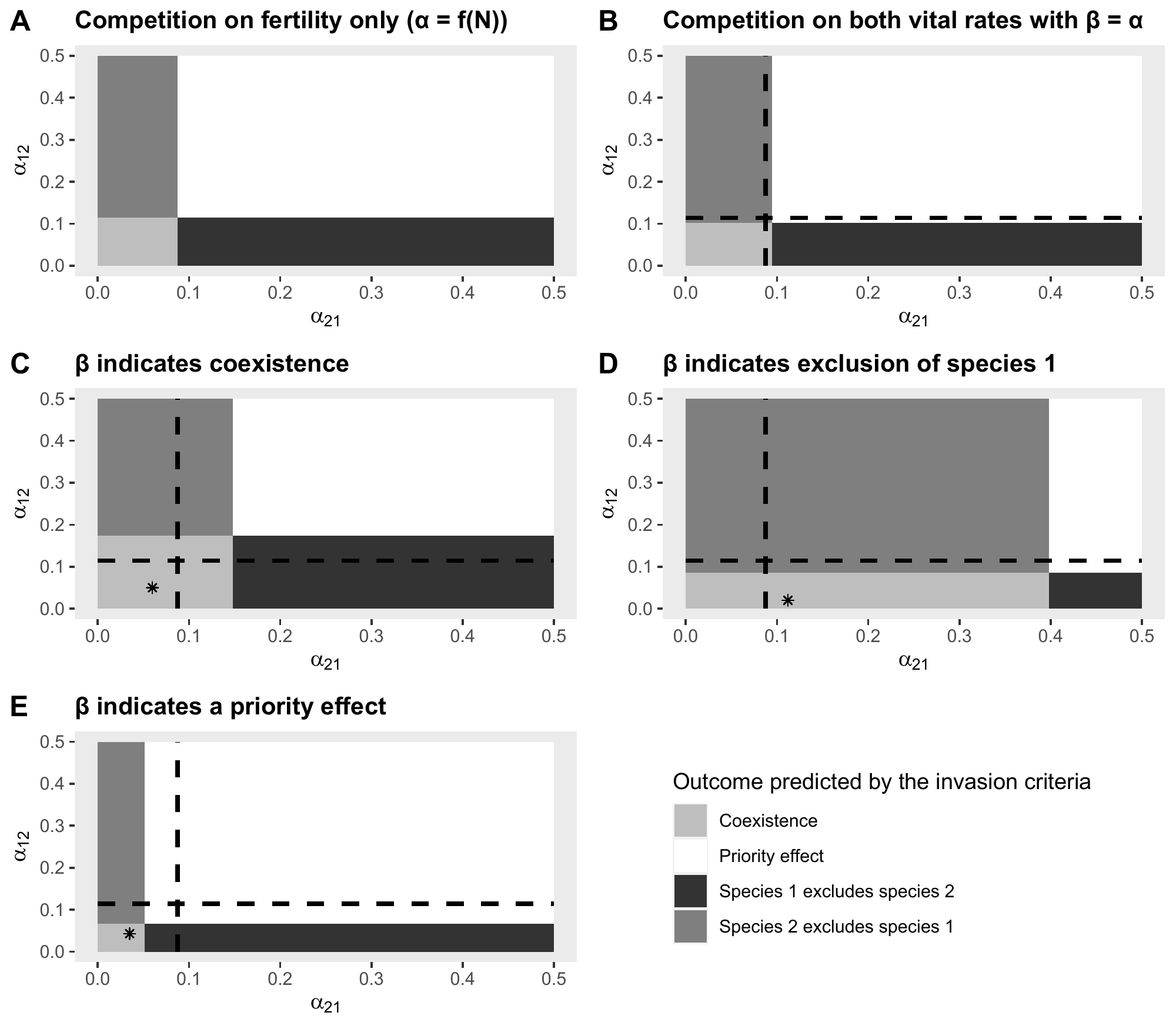}
    \caption{Outcomes of the model predicted by invasion analysis depending on the coefficients of the inter-specific competition on fertility ($\alpha_{12}$ and $\alpha_{21})$ with $\beta$ coefficients indicating coexistence (C), exclusion of species 1 (D) and a priority effect (E). The dotted lines correspond to the outcome boundaries of panel A which is used as a reference. The three asterisks stand for the locations of the three parameter sets corresponding to contrasted coexistence scenarios.}
    \label{fig:outcome_prediction}
\end{figure}

To better visualize coexistence properties implied by our invasion analysis, we produced coexistence outcome domains for model 1. These are similar to \citet{moll2008competition}'s but are computed analytically.
They are produced using the newly derived invasion criterion (eq.~\ref{eq:invasion_criteria}). Figures \ref{fig:outcome_prediction}A and \ref{fig:outcome_prediction}B are used for reference (with dotted lines corresponding to the domains boundaries of figure \ref{fig:outcome_prediction}A, which materializes what coexistence outcomes are predicted based on $\alpha$ coefficients). We found, as shown in Figs. \ref{fig:outcome_prediction}C, \ref{fig:outcome_prediction}D and \ref{fig:outcome_prediction}E, that emergent coexistence outcomes can be observed in our model. This is best seen in Fig. \ref{fig:outcome_prediction}D, where the coexistence domain in our model combining $\alpha$ and $\beta$ coefficients (in light grey) is extended far to the right of the reference dashed line, well beyond what would be indicated based on $\alpha$ coefficients only (Fig. \ref{fig:outcome_prediction}A). Thus we have coexistence with competition on both vital rates when competition on a single vital rate ($\alpha$) would instead have indicated exclusion of species 2, while $\beta$ indicates exclusion of species 1. The three parameter sets highlighted in the Models and Methods section are marked by asterisks, as panels C, D, E actually include a larger span of parameter space since $(\alpha_{12},\alpha_{21})$ do vary.

In Figs. \ref{fig:outcome_prediction}C and \ref{fig:outcome_prediction}E,
we do not find strictly emergent coexistence sensu \citet{moll2008competition}, as they restricted this label to coexistence which is not indicated by either $\alpha$ or $\beta$ but is obtained by combining both types of competition. We do find other \textit{emergent outcomes} though (e.g. in Fig. \ref{fig:outcome_prediction}E: exclusion of one of the species when $\alpha$ or $\beta$ would indicate priority effect and coexistence, respectively).

We also reproduced Fig. \ref{fig:outcome_prediction} for model 2 where habitats of juveniles and adults are separated so that they cannot compete with each other, as in the metamorphosis structure model studied by \citet{moll2008competition} (see \ref{SI:inv_analysis_MB} for details). The results are very similar.

\subsection{Sensitivity analysis}

\begin{figure}[H]
    \centering
    \includegraphics[width=0.8\linewidth]{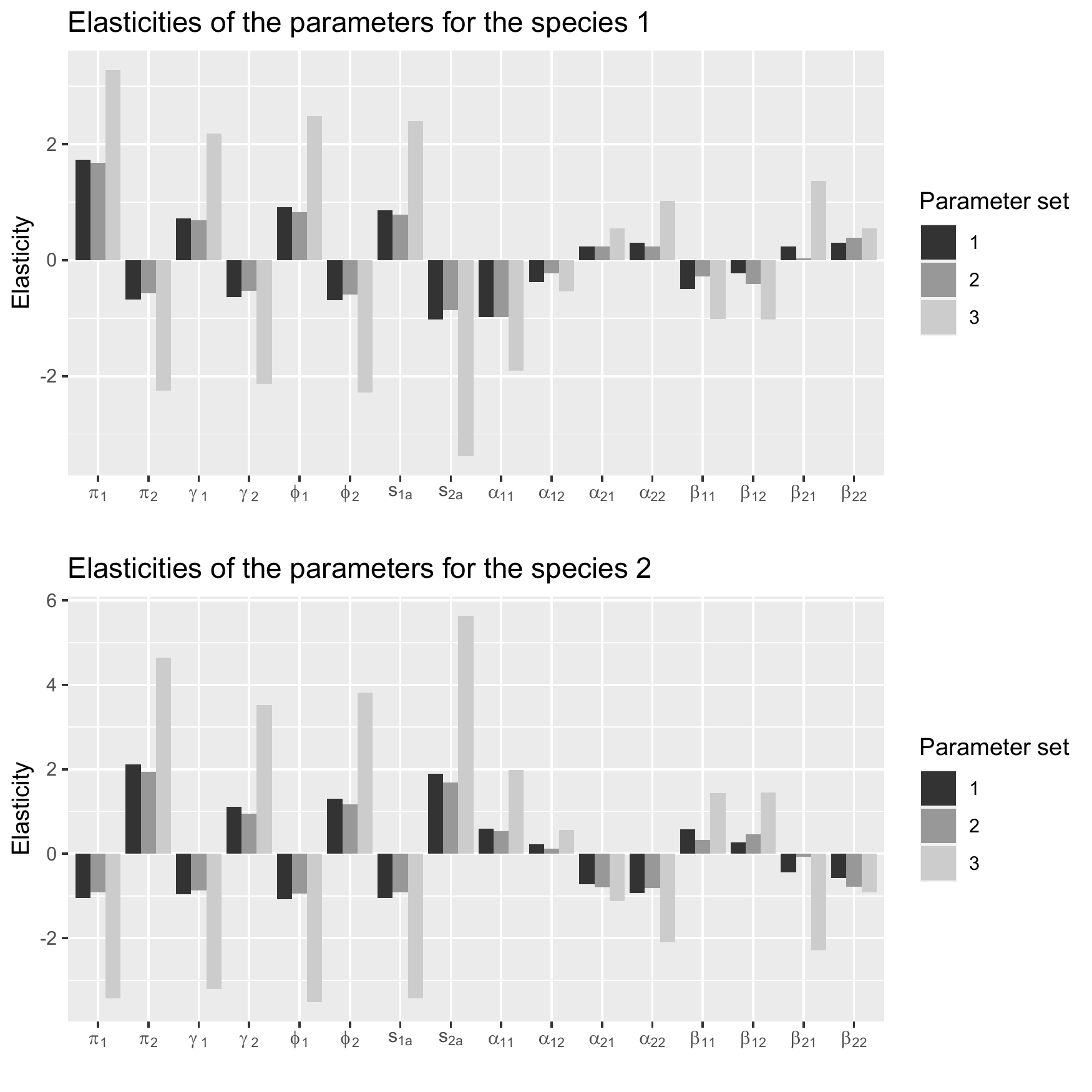}
    \caption{Elasticity of the equilibrium population
densities to changes in parameters for the 3 parameter sets of Table \ref{tab:3param_sets_inv_crit}.}
    \label{fig:tab_elast_3sets}
\end{figure}

Sensitivity and elasticity computation at the equilibrium for the three scenarios provides information relative to how much a small perturbation can change the equilibrium densities, and therefore how robust coexistence is to change in parameter values: if a small change in a parameter can make the density of a species greatly decrease, it is not robust. With some approximation (a linear one), it also provides information on how a large perturbation in parameter values could generate exclusion of one species by moving the equilibrium densities of this species to 0 (see for this the coexistence regions in \ref{app:coex_regions}). We represented elasticity $\frac{E}{n_i}\frac{dn_i}{dE}$, i.e., a relative response of the density $n_i$ of species $i$ to a relative perturbation on $E$. The elasticity values obtained from the sensitivity analysis performed on our three parameter sets are given in Fig. \ref{fig:tab_elast_3sets}. 

 We found here that the emergent coexistence that occurs when there are opposite competitive hierarchies on fertility and juvenile survival (parameter set 2) is as sensitive as the `classical' (niche-based) coexistence that occurs when the competition coefficients on both vital rates indicate coexistence (parameter set 1). However, when coexistence stems from competition on one vital rate indicating a priority effect and the other vital rate indicating niche-based coexistence (parameter set 3), the resulting equilibrium is more sensitive to perturbation, according to our elasticity analysis. Note that sensitivity/elasticity values are formally valid only close to the fixed point equilibrium. The coexistence regions representation (\citealt{barabas2014sensitivity}, presented in \ref{app:coex_regions}), which stretches a bit that validity, confirmed the results of Fig. 2 by showing similar regions for the `classical' and the `emergent' coexistence, together with a smaller domain for the third scenario (priority effect on $\beta$ + coexistence on $\alpha$ $\rightarrow$ coexistence). This is likely due to the destabilizing influence of priority effects. 

\subsection{$S$-species simulations}

\begin{figure}[H]
    \centering
    \includegraphics[scale=0.9]{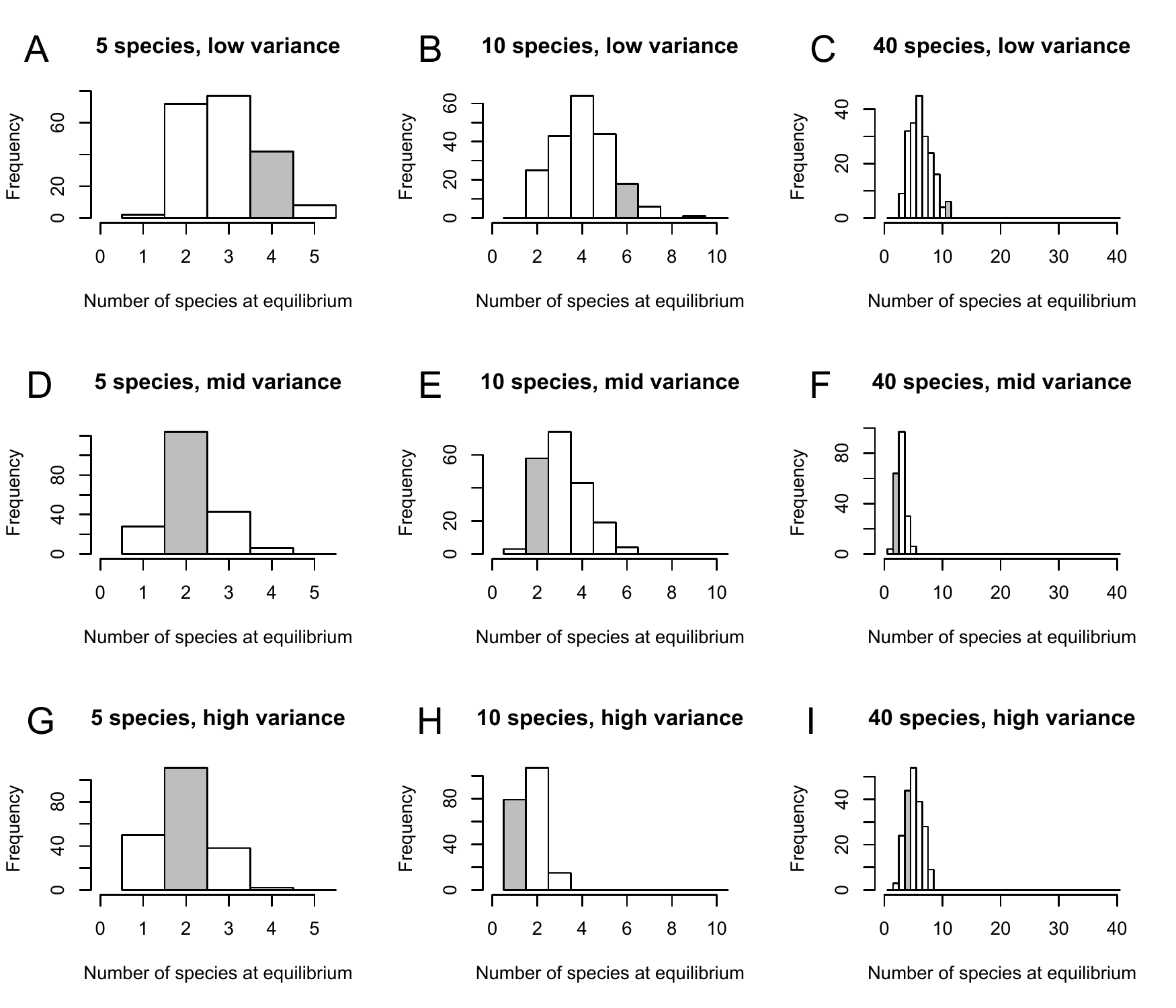}
    \caption{Histograms of the number of species persisting at equilibrium for 200 permutations of inter-specific competition coefficients. The bar in gray indicates the number of species persisting at time $t=3000$ time steps for the original $\alpha_{ij}$ and $\beta_{ij}$ with $\text{Corr}(\alpha_{ij}, \beta_{ij})<0$. From left to right, columns correspond the communities with initially 5, 10 and 40 species. From top to bottom, the lines correspond to situations with low (A-B-C), average (D-E-F) and high (G-H-I) variance on the vital rates.}
    \label{fig:hist_opp_hier_only}
\end{figure}

The species richnesses at the end of the simulations for the extended model 1 with $S$ species (see eq \ref{eq:combined_model_Sspecies}) are given in Fig. \ref{fig:hist_opp_hier_only} (see \ref{SI:sim_model_MB} for results with extended model 2). Each panel displays a histogram of the number of species remaining in the community after 3000 time steps, for the original parameter set including a negative correlation between $\alpha_{ij}$ and $\beta_{ij}$, as well as for the 200 permutated parameter sets where this correlation is set to zero.

We checked that the negative correlation between $\alpha_{ij}$ and $\beta_{ij}$ indeed induces opposite hierarchies of competition in sets of coefficients associated to fertility and juvenile survival (i.e., a negative correlation between $\text{rank}(\alpha_{ij})$ and $\text{rank}(\beta_{ij})$, see \ref{app:S_species_simus}).
In spite of this, the parameter sets with the above-mentioned opposite competitive hierarchies do not systematically lead to richer communities than the corresponding permutated parameter sets (where $\text{Corr}(\alpha_{ij}, \beta_{ij})=0$).
For very close parameters between species (low variances in parameter distributions, Figs.~\ref{fig:hist_opp_hier_only}A, \ref{fig:hist_opp_hier_only}B and \ref{fig:hist_opp_hier_only}C), opposite competitive hierarchies seem to have a positive effect on the species richness, when compared to the null hypothesis of zero correlation between $\alpha_{ij}$ and $\beta_{ij}$. However, the assumption of a community composed solely of very similar species for all life-history parameters (as opposed to similar averaged birth and death rates, as in neutral theory) is rather unrealistic.
To sum up, we showed that the negative correlation between $\alpha_{ij}$ and $\beta_{ij}$, even if it indeed induces opposite competitive hierarchies on the two vital rates, does not seem to allow, for reasonably variable distributions of parameter values, an increase of the number of species coexisting at the equilibrium. In other words, two-species coexistence through differing competitive hierarchies on life-history parameters does not scale up to many-species. 

\section{Discussion}
    
In this article, we studied competitive interactions between species with structured life-cycles, with multiple vital rates that can depend on population densities. Using structured models for two competing species, \citet{moll2008competition} described a form of coexistence whereby opposite competitive hierarchies in the competition coefficients associated to two vital rates could yield coexistence in the model where these two density-dependent vital rates combine---while models with a single density-dependent vital rate would predict exclusion of the inferior competitor. To show the existence of such emergent coexistence, they used Ricker functions for density-dependence and a particular interaction setup where competition between adults and juveniles is forbidden, as in some systems with metamorphosis, where adults and juveniles occupy separated habitats. This brought into question whether these results could be generalized to other functional forms and life cycles. We considered as a baseline an alternative and perhaps more frequent life cycle setup, where adults can affect both adults and juveniles, so that both fertilities and juvenile survival rates can depend upon adult densities of both species. We first found, using structured two-species models, that emergent coexistence can be generalized to Beverton-Holt functional forms and our new life cycle, in addition to the life cycle with metamorphosis considered by \citet{moll2008competition} (see \ref{SI:inv_analysis_MB}). We provided an in-depth explanation of how emergent coexistence can occur, using analytical invasion criteria (facilitated by the use of Beverton-Holt functions). Then, we explored the robustness of such emergent coexistence equilibria with a community-level sensitivity analysis and we found that a perturbation on parameters had not more effect on equilibrium densities in the case of emergent coexistence than in a `classical' coexistence allowed by niche separation. Finally, we numerically explored the possibility for such mechanisms of emergent coexistence to manifest in $S$-species models (with $S>>2$). We observed that opposite competitive hierarchies acting on fertility vs juvenile survival, allowed by a negative correlation between the two sets of competition coefficients, did not seem to generate substantial increases in species richness in our simulated communities. Importantly, this was true both for our model 1 where all competition is generated by adults, as well as model 2 where adults only compete with adults and juveniles with juveniles--had \citet{moll2008competition} considered more species they would have found similar results to us. \\

We now dive deeper into each of the abovementioned results. As we succeeded to reproduce the emergent coexistence of \citet{moll2008competition} with a general invasion framework and an additional, frequently observed life cycle, we think that coexistence emerging through opposite competitive hierarchies on different life history traits may be a general feature of two-species two-life-stage models. We found nothing in complex life histories with habitat/niche shifts that specifically promotes this effect of stage structure on coexistence. This reinforces the importance of structured models to study interactions between species and their role in shaping community dynamics \citep{miller2011thinking}, as we find that the set of life cycles where stage structure affects coexistence is considerably extended. Furthermore, by finding an analytic invasion criterion, we ensured that emergent coexistence is not dependent on specific sets of vital rate parameters but rather a general feature of two-species two-stages models, depending on competition hierarchies.  The robustness of such coexistence equilibria to perturbation of parameters has then been confirmed by the community-level sensitivity analysis, as the emergent coexistence scenario was not more sensitive than the classical coexistence scenario (where coexistence was indicated by $\alpha$ and $\beta$ coefficients separately). To our knowledge, there has been little implementation in practice of the discrete-time sensitivity analysis for coupled nonlinear systems developed by \citet{barabas2014fixed}. Here, it has allowed us to explore in more depth the properties of equilibria, as we could not easily find an expression for equilibrium densities (they are defined by the intersection of two conic sections as shown in \ref{SI:fixedpoints_twospecies}, and parameters are so interwoven in those expressions that even if analytical solutions were to be found, these might be too complex to interpret).
The robustness of the emergent coexistence equilibria to perturbations further suggests that such coexistence may be likely even when environmental stochasticity is added to two-species two-stages models. It is therefore an explanation to keep in mind when seeing coexisting pairs of species in the wild in spite of coexistence hierarchies on a single resource axis or vital rate indicating exclusion. 

However, two-species coexistence, even with many stages, can be quite different from many-species coexistence. Our study of $S$-species communities did not show a positive effect of opposite competitive hierarchies in the two sets of competition coefficients, except in communities where species were very similar in all their parameters---but these communities are probably not very lifelike. We have done additional explorations of many-species models, using slightly different correlation structures, such as negative correlations between interspecific $\alpha$ and $\beta$ coefficients within each pair of species (see \ref{app:S_species_simus}), rather than simply a negative correlation between pooled $\alpha_{ij}$ and $\beta_{ij}$ values for all pairs of species. This stronger condition on negative correlations between competitiveness on the two vital rates had no effect on the species richness of the final community. In sum, we showed that the properties promoting emergence of coexistence through stage structure in the 2-species models could hardly be extended to many-species models where all species interact together (albeit sometimes weakly). These results resonate with those of simpler, unstructured models \citep{barabas2016effect}: criteria needed to promote $S$-species coexistence with $S \gg 2$ are much more stringent than those for two species, and require a more important separation of niches (manifesting in the ratio between intra and interspecific competition coefficients).

Finally, we should keep in mind that even if our results are both robust and supported by analytical formulas, some assumptions could be debated. While we considered iteroparous life histories, \citet{cushing2007coexistence} considered semelparous life histories with more variable dynamics, which creates additional possibilities for coexistence, with 2-cycle attractors where juveniles of species 1 co-occur with adults of species 2 and vice versa. The extension of these semelparous life histories to our new life cycle would be quite interesting. More generally, \citet{cushing2007coexistence}, \citet{moll2008competition} and ourselves considered a modified Lotka-Volterra framework, with Lotka-Volterra competition coefficients modulating the effect of adult or juvenile densities on several vital rates. However, we have no mechanistic derivations for such models (a derivation through MacArthur-style consumer-resource model with separation of time scales could perhaps be attempted).
Structured mechanistic models of resource-based competition or competition for space might inform on the realism of the $\alpha_{ij}$ and $\beta_{ij}$ correlation structures considered here \citep{goldberg1991competitive, loreau1994competitive, tillman1982resource, qi2021coexistence}. That said, we expect that even with a mechanistic perspective, our main finding will likely be robust: while coexistence emerging from life-history complexity is quite easily seen in two-species pairs, it is much more difficult to obtain in many-species models. In speciose communities, some degree of niche separation is required to prevent species to interact too much with each other. This could be promoted by many weak-few strong interactions, possibly combined with some interaction network structures for each stage \citep[e.g.][]{kinlockuncovering}  or density-dependent vital rate, as the network structure is likely to be nonrandom and could differ among stages.

\paragraph{Acknowledgements}
We thank Sam Boireaud for an exploratory numerical study of some two-species models, as well as Olivier Gimenez for discussions and Coralie Picoche for comments on the manuscript. We are grateful to three reviewers for their constructive input.

\section*{Statements and Declarations}
\paragraph{Code accessibility}
Computer codes written for analyses are available at GitHub:\\
\url{https://github.com/g-bardon/StructuredModels_Coexistence}

\paragraph{Contributions}
GB wrote the computer code, produced the figures, and performed numerical analyses. Analytic formulas were derived by FB and GB. The first draft was written by GB with help from FB, both authors contributed equally to subsequent versions. 

\paragraph{Competing interests and funding}
The authors have no relevant financial or non-financial interests to disclose. Funding for Gaël Bardon's internship was provided by the French National Research Agency through ANR Democom (ANR-16-CE02-0007) to Olivier Gimenez.

\bibliography{stage_structure_coexistence.bib}
\bibliographystyle{amnat} 

\newpage

\appendix

\setcounter{equation}{0}
\renewcommand{\theequation}{A\arabic{equation}}

\setcounter{table}{0}
\renewcommand{\thetable}{A\arabic{table}}

\setcounter{figure}{0}
\renewcommand{\thefigure}{A\arabic{figure}}

\setcounter{section}{0}
\renewcommand{\thesection}{Appendix \arabic{section}}

\section{Invasion analysis}\label{app:inv_analysis}

We develop here the method to obtain invasion criteria in two-species two-life stages models including competition on two vital rates, such as model 1 combining competition on fertility and juvenile survival.

We first place ourselves at the exclusion equilibrium where species 1 lives without species 2:
\begin{equation}
    \tilde{n}= \begin{pmatrix}
    n_{1j}^*\\n_{1a}^*\\0\\0\\
    \end{pmatrix}.
\end{equation}

To compute the equilibrium densities $n_{1j}^*$ and $n_{1a}^*$, we have to solve the following system corresponding to the model with a single species (we dropped here species-specific indices for simplicity):
\begin{equation}
    \left\{
\begin{array}{r c l}
n_j &=& (1-\gamma)s_j(n)n_j + f(n)n_a\\
n_a &=& \gamma s_j(n)n_j + s_an_a
\end{array}
\right.
\end{equation}
with fertility $f(n)$ and juvenile survival $s_j(n)$ affected by  intra-specific competition only, according to the relations
\begin{equation}
    f(n)=\frac{\pi}{1+\alpha n_a} \quad  \text{and} \quad  s_j(n) = \frac{\phi}{1+\beta n_a}.
\end{equation}

We now express $n_j$ according to $n_a$:
\begin{align}
    n_j &= (1-s_a)n_a\frac{1}{\gamma s_j(n)} \\
        &= (1-s_a)\left(\frac{1+\beta n_a}{\gamma \phi}\right)n_a
\end{align}
 and then, as we have by assumption $n_a \neq 0$ and $n_j \neq 0$, we obtain
 \begin{align*}
     n_j &= (1-\gamma)s_j(n)n_j + f(n)n_a\\
     \Leftrightarrow 1 &= (1-\gamma)\frac{\phi}{1+\beta n_a} + \frac{\pi}{1+\alpha n_a}\frac{n_a}{n_j} \\ 
     \Leftrightarrow  1 &= (1-\gamma)\frac{\phi}{1+\beta n_a} + \frac{\pi \phi \gamma}{(1-s_a)(1+\beta n_a)(1+\alpha n_a)}\\
     \Leftrightarrow (1+\beta n_a)(1+\alpha n_a) &= (1-\gamma)\phi(1+\alpha n_a) + \frac{\pi \gamma \phi}{1-s_a}
 \end{align*}
that leads to the polynomial of degree 2:
\begin{equation}
    \alpha\beta n_a^2 + (\alpha + \beta - \alpha(1-\gamma)\phi)n_a+1-(1-\gamma)\phi-\frac{\pi\gamma\phi}{1-s_a} = 0.
\end{equation}
We denote $C= (1-\gamma)\phi$ and $D= \frac{\pi\gamma\phi}{1-s_a}$ and we find the solutions of the polynomial equation:
\begin{equation}
    n_a^* = \frac{(\alpha C-\alpha-\beta) \pm \sqrt{(-\alpha C + \alpha + \beta)^2 - 4\alpha\beta(-C-D+1)}}{2\alpha\beta}.
\end{equation}
We search for a positive solution and we have $C=(1-\gamma)\phi < 1$ which implies $\alpha C - \alpha - \beta < 0$ assuming positive interaction coefficients ($\alpha>0$ and $\beta>0$). Positivity is then equivalent to:
\begin{align}
    (\alpha C-\alpha-\beta) + \sqrt{(-\alpha C + \alpha + \beta)^2 - 4\alpha\beta(-C-D+1)} > 0 \\
    \iff - 4\alpha\beta(-C-D+1) > 0 \\
    \iff C+D > 1.
\end{align}

The equivalence is provided by the fact that the left term has its square inside the square root. We define the inherent net reproductive number as $C+D = (1-\gamma)\phi +\frac{\pi \gamma \phi}{1-s_a}$ that must exceed 1 to have a viable species. We finally obtain the densities at a stable positive equilibrium for the model with a single species:
\begin{equation}
    \left\{
\begin{array}{r c l}
n_j^* &=&  (1-s_a)(\frac{1+\beta n_a^*}{\gamma \phi})n_a^*\\
n_a^* &=& \frac{(\alpha (1-\gamma)\phi-\alpha-\beta) + \sqrt{(-\alpha (1-\gamma)\phi + \alpha + \beta)^2 - 4\alpha\beta(-(1-\gamma)\phi-\frac{\pi\gamma\phi}{1-s_a}+1)}}{2\alpha\beta}
\end{array}
\right. \label{eq:fixedpointeq_model1_singlesp}
\end{equation}
provided that
\begin{equation}
   C+D = (1-\gamma)\phi +\frac{\pi \gamma \phi}{1-s_a} > 1. 
\end{equation}
We have then obtained the densities at the exclusion equilibrium. We have now to evaluate the stability of the equilibrium $(n_j^*,n_a^*,0,0)$: if it is locally asymptotically stable, the excluded species cannot invade the community starting from very small density. Conversely, if it is unstable, this means that the excluded species can invade \textit{provided that equilibria of the single-species model are restricted to stable fixed points}. This last restriction is important. Indeed, if the single-species model has a stable two-cycle or an even more complex fluctuating attractor, the fixed point will already be unstable, and a proper invasion analysis should instead be done on this more complex attractor. Two-cycles tend to occur for $s_a=0$ \citep{cushing2007coexistence}, as shown in \ref{SI:simuls_twospecies}. So long as $s_a>0$ this model does not show 2-cycles---so far as we have seen numerically and following the argumentation of \citet{cushing2007coexistence} for model 2. The combination $s_a = 0$ and $\gamma = 1$ renders the projection matrix primitive, which in turns creates possibilities for bifurcations towards two-cycles directly from the trivial single-species (0,0) equilibrium \citep{cushing1989ebenman,cushing2006nonlinear}. We therefore restrict ourselve to the case $s_a>0$, although the equations derived above in this Appendix for the single-species fixed point are also valid whenever $s_a=0$. 

To evaluate the (un)stability of the exclusion equilibria in the two-species system, we have to compute the eigenvalues of the Jacobian of the 2 species system evaluated at the exclusion equilibrium. The full system iterated over a time step reads:
\begin{equation}
    \left\{
\begin{array}{r c l}
n_{1j}(t+1) &=& (1-\gamma_1)\frac{\phi_1}{1+\beta_{11}n_{1a}(t)+\beta_{12}n_{2a}(t)}n_{1j}(t) + \frac{\pi_1}{1+\alpha_{11}n_{1a}(t)+\alpha_{12}n_{2a}(t)}n_{1a}(t)\\
n_{1a}(t+1) &=& \gamma_1 \frac{\phi_1}{1+\beta_{11}n_{1a}(t)+\beta_{12}n_{2a}(t)}n_{1j}(t) + s_{1a}n_{1a}(t)\\
n_{2j}(t+1) &=& (1-\gamma_2)\frac{\phi_2}{1+\beta_{22}n_{2a}(t)+\beta_{21}n_{1a}(t)}n_{2j}(t) + \frac{\pi_2}{1+\alpha_{22}n_{2a}(t)+\alpha_{21}n_{1a}(t)}n_{2a}(t)\\
n_{2a}(t+1) &=& \gamma_2 \frac{\phi_2}{1+\beta_{22}n_{2a}(t)+\beta_{21}n_{1a}(t)}n_{2j}(t) + s_{2a}n_{2a}(t).
\end{array}
\right.
\end{equation}

We place ourselves at the abovementioned exclusion equilibrium where species 1 dominates the community and species 2 is absent. The Jacobian evaluated at this point is a $4 \times 4$ matrix and it has the following triangular block form (see also \citealt{cushing2008matrix}):
\begin{equation}
    \begin{pmatrix}
    B_1 & B_2 \\
    0 & B_3 \\
    \end{pmatrix}.
\end{equation}
Therefore, we only need to know the eigenvalues of the $2 \times 2$ matrices $B_1$ and $B_3$. 
The $B_1$ matrix corresponds to the Jacobian of the system in the absence of species 2, so provided that the single-species equilibrium is stable, we can infer that it has a spectral radius lower than 1. Unfortunately, although all simulations that we have done, even for very large fertilities or low competition coefficients, provide a stable equilibrium for $C+D>1$, it is not straightforward to show local stability analytically. Indeed matrix $B_1$ writes:

\begin{equation}
    \begin{pmatrix}
        (1-\gamma) \frac{\phi}{1+\beta n_a} & \frac{1-\gamma}{\gamma} \left( \frac{-\beta n_a}{1+\beta n_a}\right) (1-s_a) + \frac{\pi}{1+\alpha n_a} \left( 1-\frac{\alpha n_a}{1+\alpha n_a}\right)\\
        \gamma \frac{\phi}{1+\beta n_a} & \frac{-\beta n_a}{1+\beta n_a}+s_a
    \end{pmatrix}
\end{equation}

with $n_a$ given by eq. \eqref{eq:fixedpointeq_model1_singlesp}, making the use of Jury conditions (used below on the simpler $B_3$ block) difficult here, although the determinant simplifies somewhat. We therefore conjecture the stability of the single-species equilibrium, with a spectral radius of $B_1$ below 1, which we deem reasonable in this context.
The $B_3$ matrix is given by:
\begin{equation}
    B_3 =
    \begin{pmatrix}
    (1-\gamma_2)\frac{\phi_2}{1+\beta_{21}n_{1a}^*} & \frac{\pi_2}{1+\alpha_{21}n_{1a}^*} \\
    \gamma_2\frac{\phi_2}{1+\beta_{21}n_{1a}^*} & s_{2a} \\
    \end{pmatrix}.
\end{equation}
The $B_3$ matrix block, which is part of the Jacobian, also corresponds to the projection matrix of species 2 when invading species 1, which implies that the exclusion equilibrium is not stable when species 2 has a growth rate larger than 1 (i.e., the modulus of $B_3$'s leading eigenvalue is larger than 1). The stability of this two-dimensional system can be investigated thanks to the Jury conditions, given by the equations (\ref{SI1st_jury_condition}, \ref{SI2nd_jury_condition}, \ref{SI3rd_jury_condition}):
\begin{align}
1- \tr J(\mathbf{n})+\det J(\mathbf{n}) &>0\label{SI1st_jury_condition}\\ 
1+ \tr J(\mathbf{n}) + \det J(\mathbf{n}) &> 0 \label{SI2nd_jury_condition}\\
1- \det J(\mathbf{n}) &> 0 \label{SI3rd_jury_condition}
\end{align}
with $J(\mathbf{n})$ the Jacobian matrix of the two-compartment system whose stability is being evaluated. If these three conditions hold, the system is stable. If the first condition (equation \eqref{SI1st_jury_condition}) is violated, one of the eigenvalues of $J(\mathbf{n})$ is larger than 1, which means in our context that the excluded species can invade the community. The violations of the conditions \eqref{SI2nd_jury_condition} and \eqref{SI3rd_jury_condition} correspond to the creation of limit cycles or invariant loops \citep{neubert2000density}. We then check these conditions on the $B_3$ matrix. We have:
\begin{align*}
    \tr B_3 &= (1-\gamma_2)\frac{\phi_2}{1+\beta_{21}n_{1a}^*} + s_{2a} \\
    \det B_3 &= (1-\gamma_2)\frac{\phi_2}{1+\beta_{21}n_{1a}^*}s_{2a} - \gamma_2\frac{\phi_2}{1+\beta_{21}n_{1a}^*}\frac{\pi_2}{1+\alpha_{21}n_{1a}^*}.
\end{align*}
We start with the third condition (\ref{SI3rd_jury_condition}):
\begin{equation}\label{eq:SIJury3}
    1- (1-\gamma_2)\frac{\phi_2}{1+\beta_{21}n_{1a}^*}s_{2a} + \gamma_2\frac{\phi_2}{1+\beta_{21}n_{1a}^*}\frac{\pi_2}{1+\alpha_{21}n_{1a}^*} > 0.
\end{equation}
This equation will always be satisfied with biologically meaningful parameters and strictly competitive interaction between species: 
\begin{itemize}
    \item The third term of \eqref{eq:SIJury3} is positive because each parameter is positive,
    \item $(1-\gamma_2)\frac{\phi_2}{1+\beta_{21}n_{1a}^*}s_{2a} < 1$ because $0 \leq \gamma \leq 1$, $0 \leq \phi \leq 1$ and all $\beta$'s are positive.
\end{itemize}
Since $\tr B_3 > 0$, the first Jury's condition implies the second ((\ref{1st_jury_condition}) $\Rightarrow$ (\ref{2nd_jury_condition})). Then, we only have to check the first Jury condition to describe the stability of the system:
\begin{align*}
    1-(1-\gamma_2)\frac{\phi_2}{1+\beta_{21}n_{1a}^*} + s_{2a}+(1-\gamma_2)\frac{\phi_2}{1+\beta_{21}n_{1a}^*}s_{2a} - \gamma_2\frac{\phi_2}{1+\beta_{21}n_{1a}^*}\frac{\pi_2}{1+\alpha_{21}n_{1a}^*} > 0 \\
    (1-\gamma_2)\frac{\phi_2}{1+\beta_{21}n_{1a}^*} - (1-\gamma_2)\frac{\phi_2}{1+\beta_{21}n_{1a}^*}s_{2a} + \gamma_2\frac{\phi_2}{1+\beta_{21}n_{1a}^*}\frac{\pi_2}{1+\alpha_{21}n_{1a}^*} < 1-s_{2a}\\
    (1-\gamma_2)\phi_2(1-s_{2a}) < (1-s_{2a})(1+\beta_{21}n_{1a}^*)- \frac{\gamma_2\phi_2\pi_2}{1+\alpha_{21}n_{1a}^*}\\
    (1-\gamma_2)\phi_2\frac{1}{1+\beta_{21}n_{1a}^*} + \frac{\gamma_2\phi_2\pi_2}{1-s_{2a}}\frac{1}{(1+\beta_{21}n_{1a}^*)(1+\alpha_{21}n_{1a}^*)} < 1.
\end{align*}
We finally have the following condition for the stability of the exclusion equilibrium:
\begin{equation}
    \frac{C_2}{1+\beta_{21}n_{1a}^*}+\frac{D_2}{(1+\beta_{21}n_{1a}^*)(1+\alpha_{21}n_{1a}^*)} < 1
\end{equation}
with $C_2= (1-\gamma_2)\phi_2$ and $D_2= \frac{\pi_2\gamma_2\phi_2}{1-s_{2a}}$.\\

This last expression (eq.~\ref{eq:invasion_criteria}) gives us an invasion criteria that is larger than 1 if, conversely, species 2 is able to invade the community when rare.

Note that if we remove the competition on one vital rate by setting all the competition coefficients associated to this vital rate to 0, we re-obtain the invasion criteria given by \citet{fujiwara2011coexistence} for their model where a single vital rate was affected by competition. This ensures some internal coherence to the analytical results. 

\section{Coexistence regions obtained through sensitivity analysis for the three scenarios of coexistence}\label{app:coex_regions}

\begin{figure}[H]
    \centering
    \includegraphics[width=\textwidth]{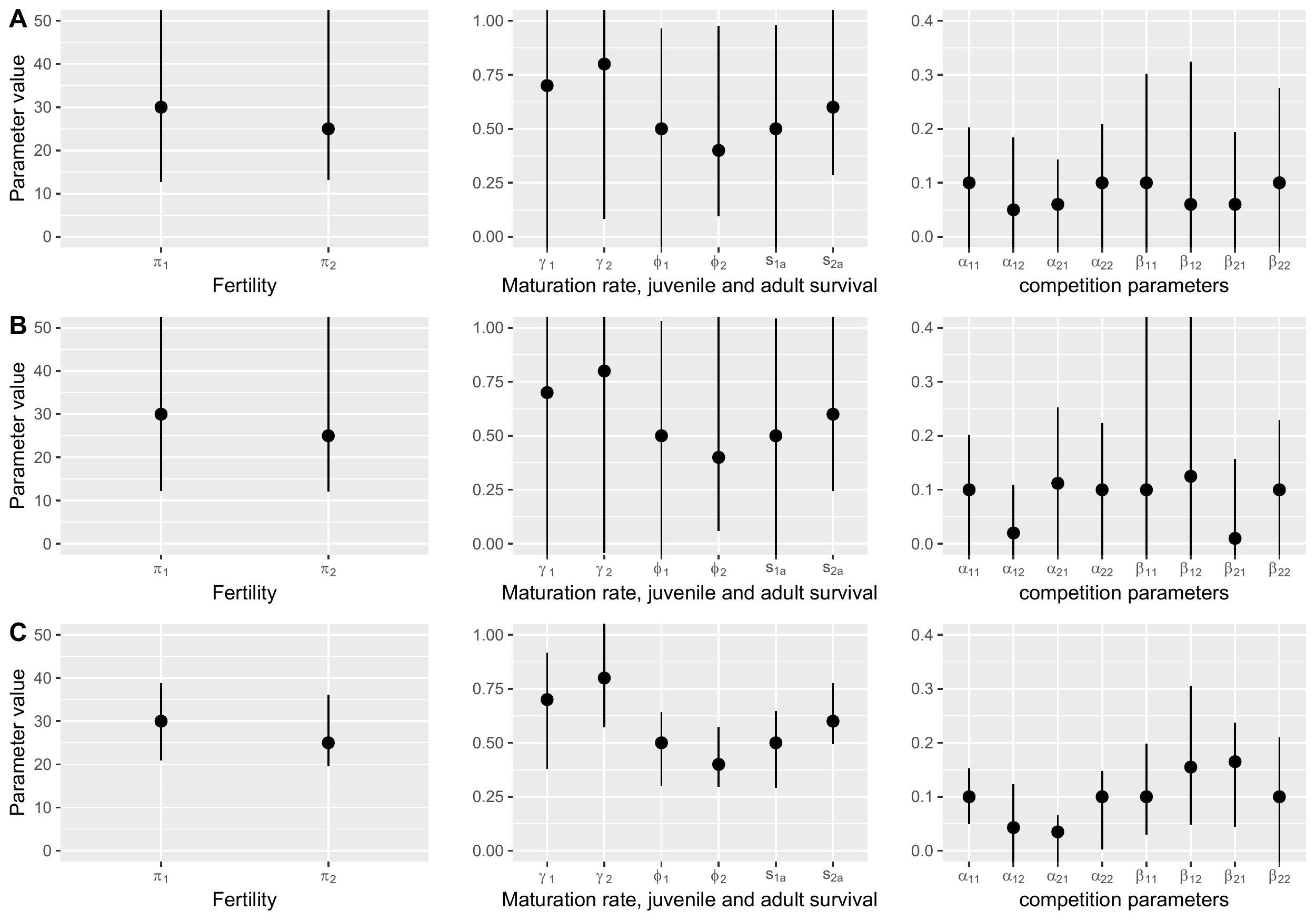}
    \caption{Parameter values and coexistence regions of the three parameter sets of Table \ref{tab:3param_sets_inv_crit} (A - set 1, B - set 2, C - set 3) chosen to promote different scenarios of coexistence.}
    \label{fig:coexistence_regions_3sets}
\end{figure}

To have another, complementary representation of the sensitivities, we estimated coexistence regions for the three parameter sets using the definition given by \citet{barabas2014sensitivity}. Note that we arbitrarily use the term \text{region} in place of \textit{domain} to separate the \textit{regions} found with this sensitivity analysis from the \textit{domains} of coexistence found through invasion analysis. 
The method consists in using the sensitivities at the equilibrium for each parameter in order to determine the region of parameter space where species both persists despite changes in parameter values. It is calculated by finding the smallest (positive and negative) perturbation that would lead to the extinction of one of the species from the value of sensitivities and densities at equilibrium, and makes the (crude) approximation that sensitivities are not only valid close to the fixed point but also far away from it.
The coexistence regions for the three parameter sets are presented in Fig. \ref{fig:coexistence_regions_3sets}.

We see here smaller coexistence regions in the third scenario, consistent with this equilibrium being more sensitive (less robust) to perturbations.

\section{Additional simulations with $S$ species, $S>2$}\label{app:S_species_simus}

The meta-parameter sets, used in the parameter distributions from which the vital rate parameters for each species have been drawn, are provided in Table \ref{tab:parameter_law_community_simul}.

\begin{table}[H]
    \centering
\begin{tabular}{|c|cc|cc|cc|cc|cc|}
  \hline
Distribution variance & $\mu_{\pi_i}$ & $\sigma_{\pi_i}$ & a & b & $\mu_{\alpha_{ii}}$ & $\sigma_{\alpha_{ii}}$ & $\mu_{\beta_{ii}}$ & $\sigma_{\beta_{ii}}$ & $\mu_{ij}$ & $\sigma^2$\\ 
  \hline
high & 3 & 0.5 & 5 & 5 & 0.05 & 0.01 & 0.05 & 0.01 & 0.04 & 0.02\\ 
average & 3 & 0.05 & 50 & 50 & 0.05 & 0.01 & 0.05 & 0.01 & 0.04 & 0.02\\ 
small & 3 & 0.005 & 500 & 500 & 0.05 & 0.001 & 0.05 & 0.001 & 0.04 & 0.02\\ 
   \hline
\end{tabular}

\caption{(Meta)-Parameter sets used to draw parameter values to simulate community dynamics with $(a,b)$ parameters of the Beta distribution for $\phi$, $s_{a}$ and $\gamma$ ($a = a_{\phi_i} = a_{s_{a,i}} = a_{\gamma_i}$ and $b= b_{\phi_i} = b_{s_{a,i}} = b_{\gamma_i}$). The last two columns correspond respectively to the mean of inter-specific coefficients (for both $\alpha_{ij}$ and $\beta_{ij}$) and to the variance of the inter-specific coefficients (with the covariance between $\alpha_{ij}$ and $\beta_{ij}$ equal to $-0.9 \times \sigma^2$).}
\label{tab:parameter_law_community_simul}
\end{table}

\begin{figure}[H]
    \centering
    \includegraphics[width=\textwidth]{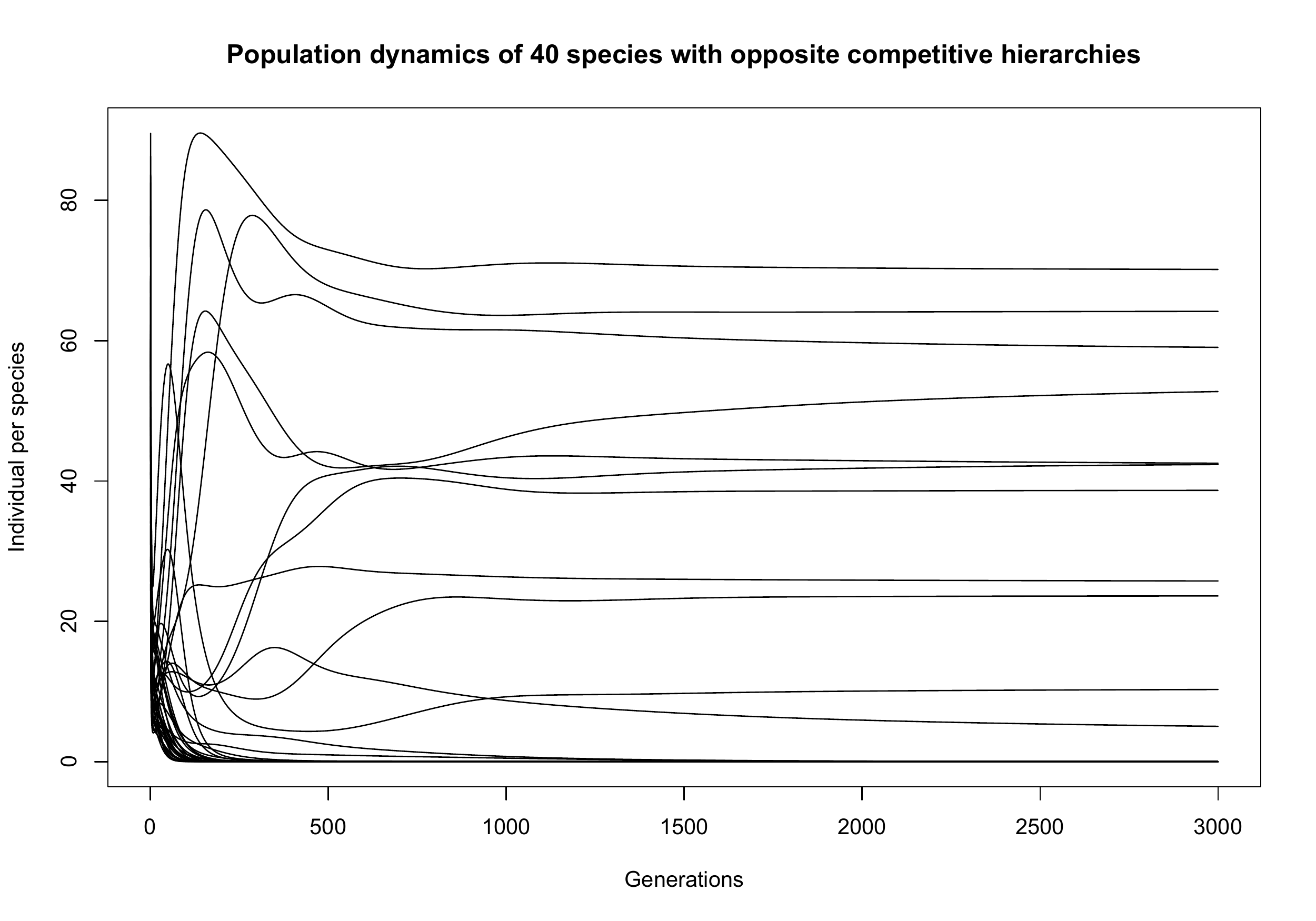}
    \caption{Population dynamics of a community with initially 40 species and opposite competitive hierarchies on fertility and juvenile survival. All species have very similar vital rates due to a low variance in the distribution used to draw the parameter. Each line corresponds to one species.}
    \label{fig:dynamic_community}
\end{figure}

For initially 40 species, a small variance of parameters across species, and opposite competitive hierarchies, we obtained the dynamics illustrated in Fig. \ref{fig:dynamic_community}.

The population dynamics highlighted in Fig. \ref{fig:dynamic_community} demonstrates that 3000 time steps are enough to reach a stable number of coexisting species.

\begin{figure}[H]
    \centering
    \includegraphics[scale=0.8]{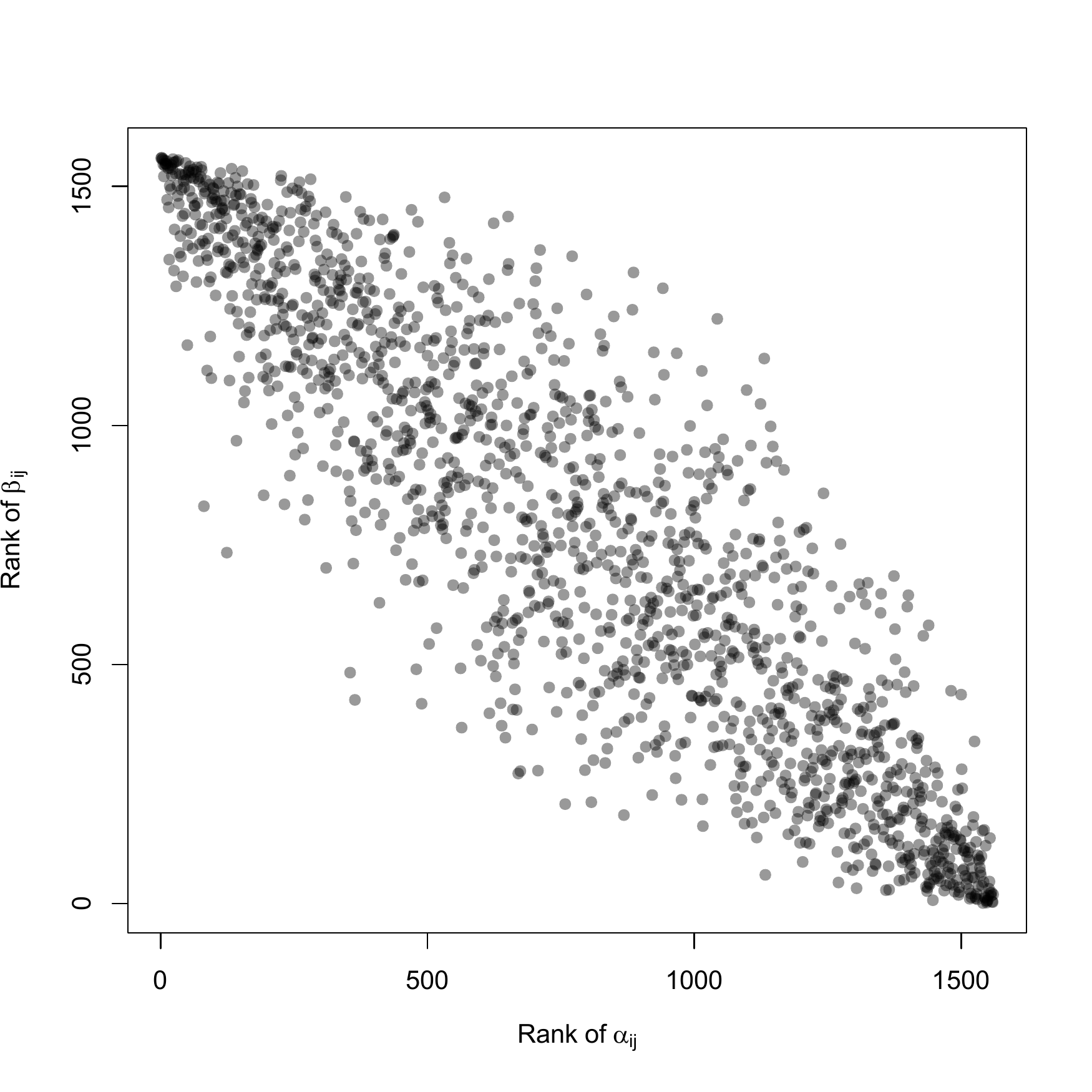}
    \caption{Rank of $\beta_{ij}$ versus rank of $\alpha_{ij}$ under a negative correlation of $\beta_{ij}$ and $\alpha_{ij}$ of  $-0.9$.}
    \label{fig:opposite_hier_alpha_beta}
\end{figure}

We verified our hypothesis that a negative correlation between $\alpha_{ij}$ and $\beta_{ij}$ lead to opposite hierarchies in sets of coefficients associated to fertility and juvenile survival by plotting the rank of $\alpha_{ij}$ against the rank of $\beta_{ij}$.
For 40 species, these ranks are plotted in Fig. \ref{fig:opposite_hier_alpha_beta}.
The negative correlation between ranks in Fig. \ref{fig:opposite_hier_alpha_beta} confirmed that the negative correlation between $\alpha_{ij}$ and $\beta_{ij}$ was creating the expected opposite competitive hierarchies. 

\begin{figure}[H]
    \centering
    \includegraphics[scale=0.8]{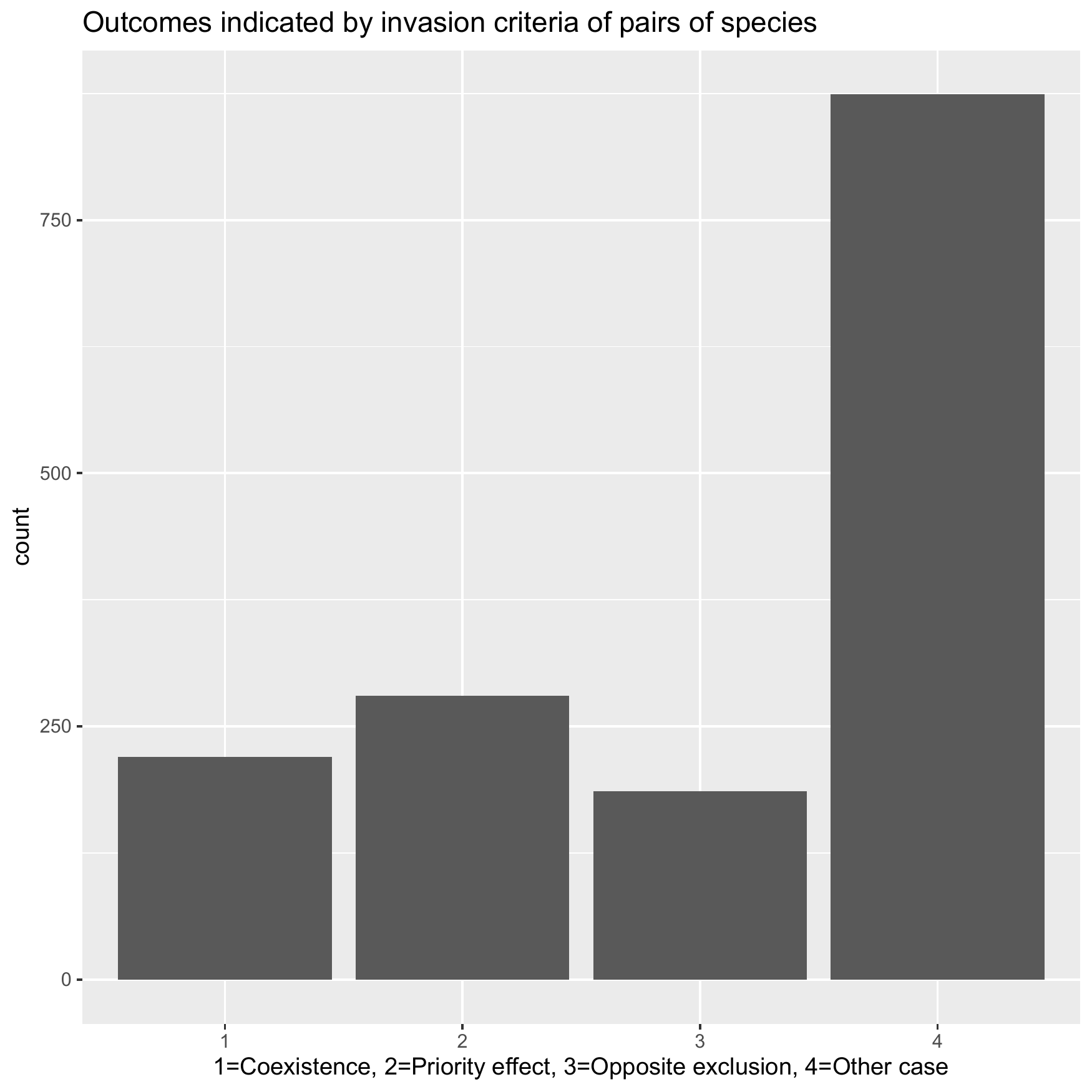}
    \caption{Frequencies of outcomes predicted by pairwise invasion criteria for a parameter set with 40 species with very close parameters and a negative correlation between $\alpha_{ij}$ and $\beta_{ij}$ with $\phi = 0.9\sigma$ }
    \label{fig:pairwise_interaction_opp_hier_only}
\end{figure}

However, opposite competitive hierarchies may not be sufficient to promote reciprocal exclusion (e.g., exclusion of 1 by 2 in the fertility competition model and of 2 by 1 in the juvenile survival competition model) of species within each or most pairs of species. To make the comparison between 2 and S-species model, we have to check that we consider situations that are fully comparable, i.e., where species would exclude each other out in a two-species contest on either $\alpha$ or $\beta$ competition. We therefore checked if the trade-off between being competitive on $\alpha$ and $\beta$ could promote such reciprocal exclusion or priority effects (in models with a single parameter that is density dependent) within pairs of species. We computed the invasion criteria of such simple models (where only one vital rate is affected by competition) for each pair of species composing the communities. For a community of initially 40 species and a small variance in parameters across species, we found the coexistence outcomes in simpler models highlighted in Fig. \ref{fig:pairwise_interaction_opp_hier_only}. 

We found that for a community of 40 species, the proportions of pairs of species where a priority effect or a reciprocal exclusion are indicated (by the simpler models with density-dependence on a single vital rate) are low when we assume only a negative correlation between $\alpha_{ij}$ and $\beta_{ij}$. 
In fact, when we simply use a negative correlation to create opposite competitive hierarchies, we draw our couple of values ($\alpha_{ij}$, $\beta_{ij}$) in a Gaussian distribution, and then most of the values are close to their means, which does not allow to obtain pairs of species where each species is strongly competitive on a different vital rate, and exclude the other if the competition was on this vital rate only. Therefore, even if there are opposite competitive hierarchies within pairs of species modelled, this setup is unlikely to generate situations where species exclude each other when considering single-vital-rate-competition. 

To consider S-species scenarios where species can exclude each other out if competition was solely on $\alpha$ or solely on $\beta$ (hereafter referred to as `reciprocal exclusion'), we changed our structure of correlation to generate situations where within most pairs of species, each species is strongly competitive on a different vital rate (e.g. species 1 excludes species 2 in the model with only fertility competition and vice-versa in the model with only juvenile survival competition). 

Our method to achieve reciprocal exclusion of pairs (for single-vital-rate-competition) consisted in drawing pairs $(a_{ij},b_{ij})$ from a bivariate normal distribution with $\mu_{a_{ij}}$<$\mu_{b_{ij}}$ and $\text{Corr}(a_{ij},b_{ij}) < 0$. We then always assign $(a_{ij},b_{ij})$ to $(\alpha_{ij}, \beta_{ij})$ and $(b_{ji},a_{ji})$ to $(\alpha_{ji}, \beta_{ji})$. In this way, for each pair of species $(i,j)$, species $i$ is competitive on $\alpha$ and species $j$ is competitive on $\beta$. We acknowledge that this may be hard to grasp, and suggest to the reader to use the code to create an example. 

\begin{figure}[H]
    \centering
    \includegraphics[scale=0.8]{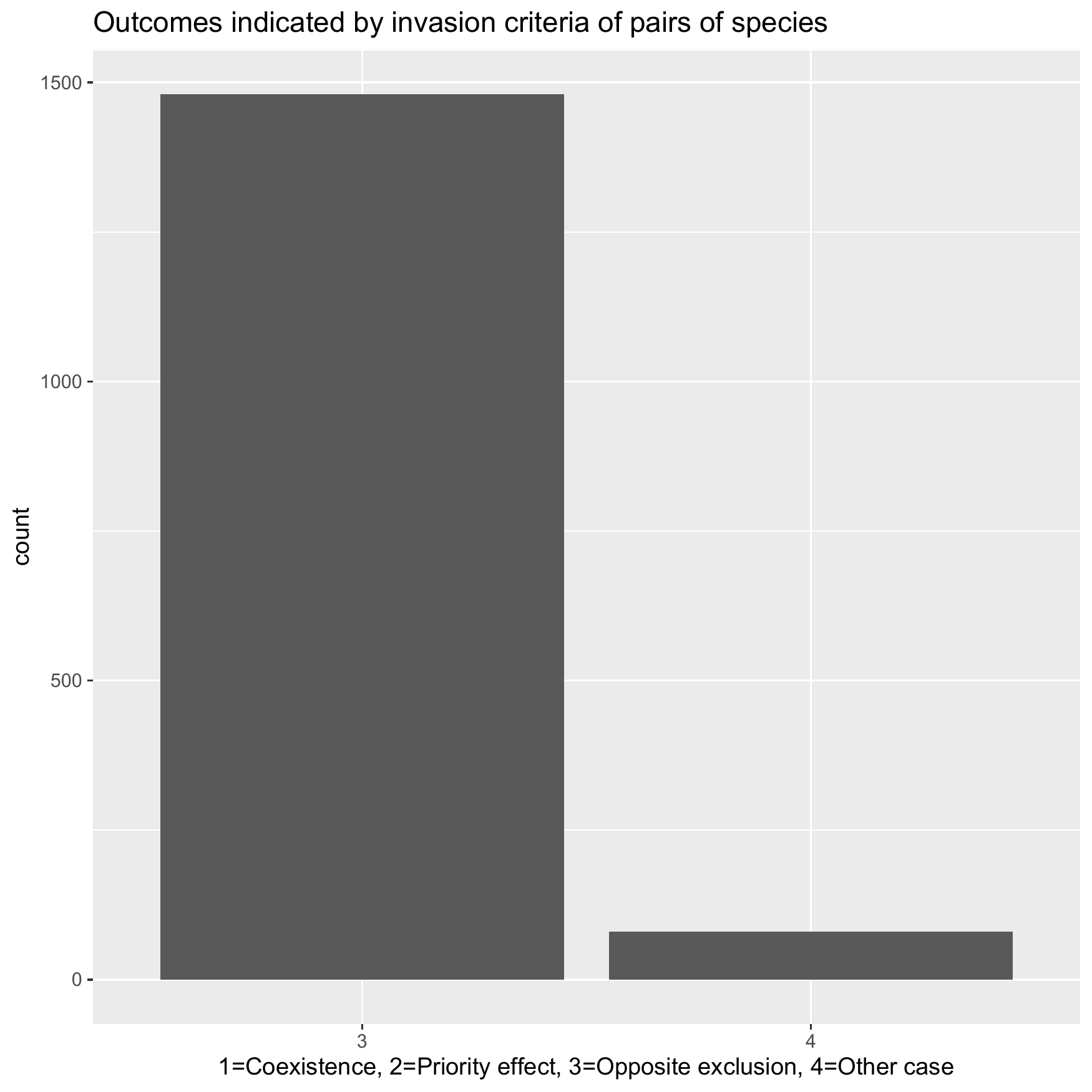}
    \caption{Frequencies of outcomes predicted by pairwise invasion criteria - for singe-vital-rate-competition models - for a parameter set with 40 species with very close parameters and bimodal distribution of interspecific coefficients drawn to promote situations of opposite exclusion within each pair of species.}
    \label{fig:pairwise_interaction_opposite_exclusion}
\end{figure}

\begin{figure}[H]
    \centering
    \includegraphics[scale=0.8]{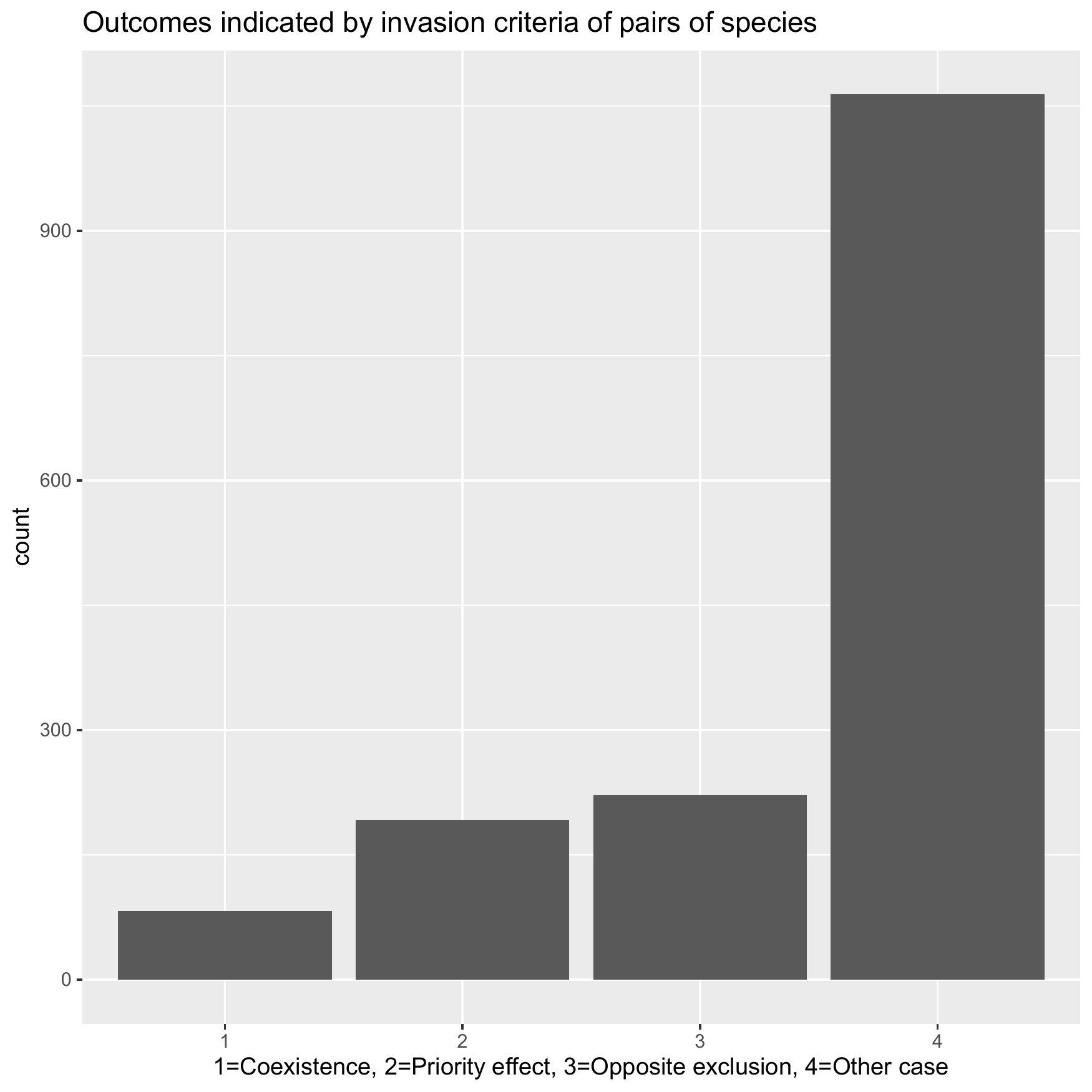}
    \caption{Frequencies of outcomes predicted by pairwise invasion criteria for a parameter set with 40 species, with very close parameters and permuted competition coefficient that promoted before permutation situations of reciprocal exclusion within each pair of species.}
    \label{fig:pairwise_interaction_opposite_exclusion_permuted}
\end{figure}

We computed the invasion criteria for each pair of species and counted how often each invasion scenario in single-vital-rate-competition models, which we represented in the histogram of Fig. \ref{fig:pairwise_interaction_opposite_exclusion}.
We succeeded to generate reciprocal exclusion within most pairs of species. Moreover, we showed that a permutation of competition coefficient was sufficient to remove the structure of reciprocal exclusion (Fig. \ref{fig:pairwise_interaction_opposite_exclusion_permuted}).

\begin{figure}[H]
    \centering
    \includegraphics[scale=0.8]{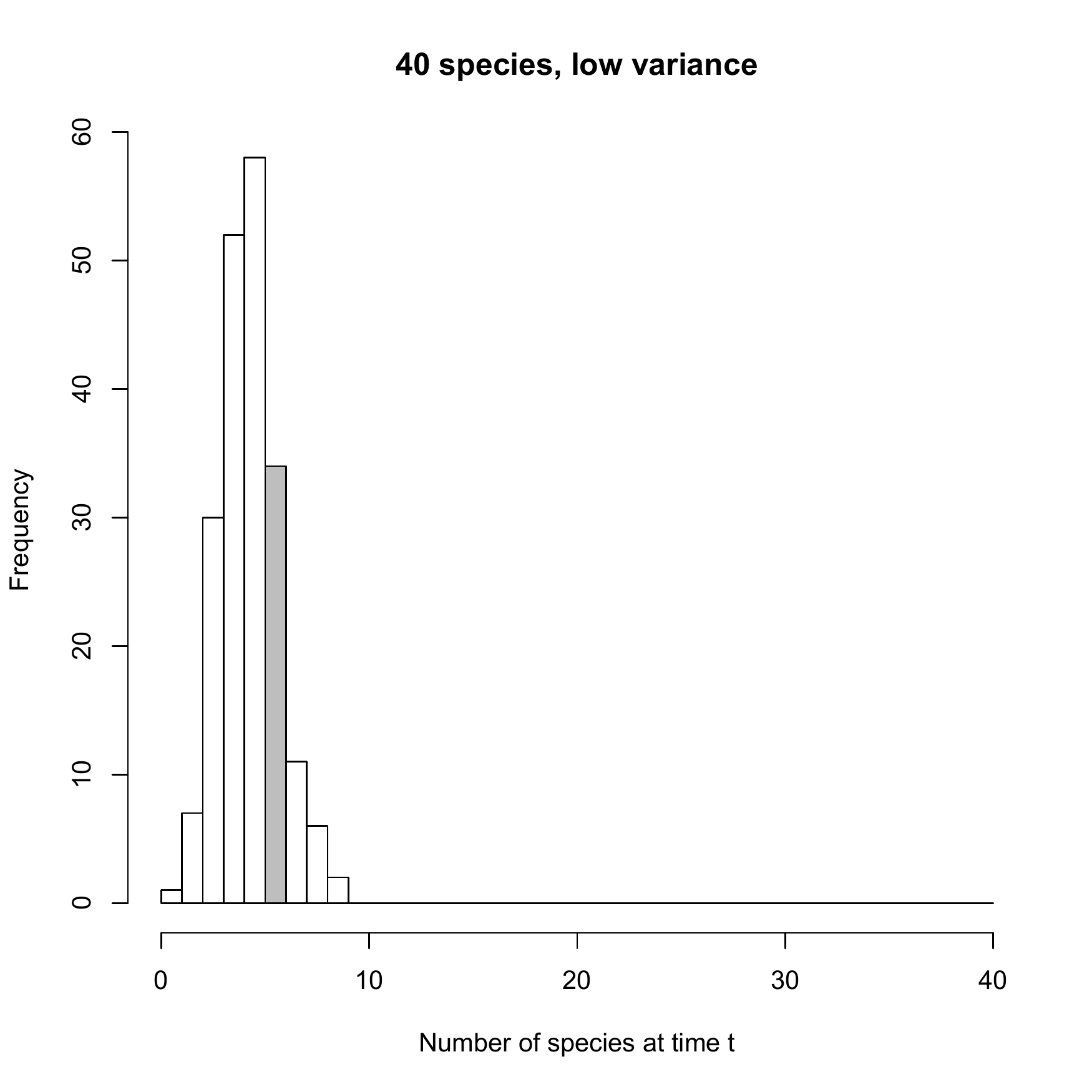}
    \caption{Histograms of the number of species persisting at equilibrium, with initially 40 species, and for 200 permutations of inter-specific competition coefficients. The bar in gray indicates the number of species persisting at equilibrium for the original $\alpha_{ij}$ and $\beta_{ij}$ with $\text{Corr}(\alpha_{ij}, \beta_{ij})<0$ and reciprocal exclusion for each pair of species.}
    \label{fig:hist_opposite_exclusion}
\end{figure}

However, as for our simple negative correlation between competitive ranks for $\alpha$ and $\beta$, this new parameter set where reciprocal exclusion occurs within most pairs of species does not allow to significantly increase the number of extant species at $t=3000$ time steps (Fig. \ref{fig:hist_opposite_exclusion}).
Therefore, even if competition is highly structured in a way that makes species exclude each other out when a single vital rate is density-dependent, which should greatly promote situations of emergent coexistence, we did not observe a positive effect of this structure on species richness in a community larger than $S=2$. We therefore conclude that emergent coexistence is highly unlikely in many-species contexts. 

\newpage

\setcounter{section}{0}
\renewcommand{\thesection}{Supplement \Alph{section}}

\setcounter{equation}{0}
\renewcommand{\theequation}{S\arabic{equation}}

\setcounter{figure}{0}
\renewcommand{\thefigure}{S\arabic{figure}}

\setcounter{table}{0}
\renewcommand{\thetable}{S\arabic{table}}

\section{Fixed points for the two-species models: equations and analytical progress in a simple case}\label{SI:fixedpoints_twospecies}

To compute the densities of both species at equilibrium for model 1, we have to solve the following system:
\begin{equation}
    \left\{
\begin{array}{r c l}
n_{1j} &=& (1-\gamma_1)\frac{\phi_1}{1+\beta_{11}n_{1a}+\beta_{12}n_{2a}}n_{1j} + \frac{\pi_1}{1+\alpha_{11}n_{1a}+\alpha_{12}n_{2a}}n_{1a}\\
n_{1a} &=& \gamma_1 \frac{\phi_1}{1+\beta_{11}n_{1a}+\beta_{12}n_{2a}}n_{1j} + s_{1a}n_{1a}\\
n_{2j} &=& (1-\gamma_2)\frac{\phi_2}{1+\beta_{22}n_{2a}+\beta_{21}n_{1a}}n_{2j} + \frac{\pi_2}{1+\alpha_{22}n_{2a}+\alpha_{21}n_{1a}}n_{2a}\\
n_{2a} &=& \gamma_2 \frac{\phi_2}{1+\beta_{22}n_{2a}+\beta_{21}n_{1a}}n_{2j} + s_{2a}n_{2a}.
\end{array}
\right.
\end{equation}

We start by expressing $n_{1j}$ according to $n_{1a}$:
\begin{align*}
    n_{1j}\left(1-\frac{(1-\gamma_1)\phi_1}{1+\beta_{11}n_{1a}+\beta_{12}n_{2a}}\right) = \frac{\pi_1}{1+\alpha_{11}n_{1a}+\alpha_{12}n_{2a}}\\
    n_{1j} = \frac{(1+\beta_{11}n_{1a}+\beta_{12}n_{2a})\pi_1}{(1+\alpha_{11}n_{1a}+\alpha_{12}n_{2a})(1+\beta_{11}n_{1a}+\beta_{12}n_{2a}-(1-\gamma_1)\phi_1}n_{1a}.
\end{align*}
Then, we insert the latter expression in the expression for $n_{1a}$:
\begin{align*}
    n_{1a} = \frac{\gamma_1\phi_1}{1+\beta_{11}n_{1a}+\beta_{12}n_{2a}}\frac{(1+\beta_{11}n_{1a}+\beta_{12}n_{2a})\pi_1}{(1+\alpha_{11}n_{1a}+\alpha_{12}n_{2a})(1+\beta_{11}n_{1a}+\beta_{12}n_{2a}-(1-\gamma_1)\phi_1}n_{1a} + s_{1a}n_{1a}.
\end{align*}
We place ourselves in the case where both species coexist. Therefore, $n_{1a} \neq 0$ and we have:
\begin{align*}
    (1+\alpha_{11}n_{1a}+\alpha_{12}n_{2a})(1+\beta_{11}n_{1a}+\beta_{12}n_{2a}-(1-\gamma_1)\phi_1) = \frac{\gamma_1\phi_1\pi_1}{1-s_{1a}}\\
    (1+\alpha_{11}n_{1a}+\alpha_{12}n_{2a})(1+\beta_{11}n_{1a}+\beta_{12}n_{2a}-C_1) = D_1\\
    \frac{C_1}{1+\beta_{11}n_{1a}+\beta_{12}n_{2a}} + \frac{D_1}{(1+\beta_{11}n_{1a}+\beta_{12}n_{2a})(1+\alpha_{11}n_{1a}+\alpha_{12}n_{2a})} = 1.
\end{align*}
This last expression correspond to the growth rate of the species 1 that must be equal to 1 at the equilibrium. \\
By symmetry, we have the same equation for the other species :
\begin{align*}
    (1+\alpha_{22}n_{2a}+\alpha_{21}n_{1a})(1+\beta_{22}n_{2a}+\beta_{21}n_{1a}-C_2) = D_2\\
    \frac{C_2}{1+\beta_{22}n_{2a}+\beta_{21}n_{1a}} + \frac{D_2}{(1+\beta_{22}n_{2a}+\beta_{21}n_{1a})(1+\alpha_{22}n_{2a}+\alpha_{21}n_{1a})} = 1.
\end{align*}

Finally, to compute the equilibrium densities, we have to solve the following system for $n_{1a}$ and $n_{2a}$ :
\begin{equation}
    \left\{
\begin{array}{r c l}
(1+\alpha_{11}n_{1a}+\alpha_{12}n_{2a})(1+\beta_{11}n_{1a}+\beta_{12}n_{2a}-C_1) = D_1\\
(1+\alpha_{22}n_{2a}+\alpha_{21}n_{1a})(1+\beta_{22}n_{2a}+\beta_{21}n_{1a}-C_2) = D_2
\end{array}
\right.
\end{equation}
which can also be written in the following form with $x=n_{1a}$ and $y=n_{2a}$ :
\begin{equation}
    \left\{
\begin{array}{r c l}
\alpha_{11}\beta_{11}x^2 + \alpha_{12}\beta_{12}y^2 + (\alpha_{11}\beta_{12}+\alpha_{12}\beta_{11})xy + (\beta_{11}+\alpha_{11}-C_1\alpha_{11})x+(\beta_{12}+\alpha_{12}-C_1\alpha_{12})y + (1-C_1-D_1) = 0\\
\alpha_{22}\beta_{22}y^2 + \alpha_{21}\beta_{21}x^2 + (\alpha_{22}\beta_{21}+\alpha_{21}\beta_{22})xy + (\beta_{22}+\alpha_{22}-C_2\alpha_{22})y+(\beta_{21}+\alpha_{21}-C_2\alpha_{21})x + (1-C_2-D_2) = 0.
\end{array}
\right.
\end{equation}

This system corresponds to two bivariate quadratics equations (quadrics). Therefore, the solutions of the system are the intersection points of the two conic sections described by each equation. These conic sections are themselves the intersection of the surface of a cone with a plane. We can classify the conic sections by looking at the discriminant of the matrix representing the conic sections.

With the standard notation of a quadratic equation:
\begin{equation}
    Ax^2+Bxy+Cy^2+Dx+Ey+F=0.
\end{equation}
We use the usual matrix representation of the conic sections that have the following general form:
\begin{equation}
    \mathbf{x^T} A_Q \mathbf{x}
\end{equation}
with 
\begin{equation}
    \mathbf{x} = \begin{pmatrix} x \\ y\\ 1 \end{pmatrix}
\end{equation}
and
\begin{equation}
    A_Q = \begin{pmatrix}
    A & B/2 & D/2 \\
    B/2 & C & E/2 \\
    D/2 & E/2 & F
\end{pmatrix}
\end{equation}
the matrix of the quadratic equation.
Then, we denote the matrix of the quadratic form $A_{33}$:
\begin{equation}
    A_{33} = \begin{pmatrix} A & B/2 \\ B/2 & C \end{pmatrix}
\end{equation}
which is obtained by removing the 3rd column and the 3rd line to the $A_Q$ matrix.

With these standard notations, we know that the conic section is degenerated if $\det A_Q = 0$ and is non-degenerated otherwise. Then, if the conic section is non-degenerated, we have the following classification :
\begin{itemize}
    \item the section is a hyperbola if and only if $\det A_{33} <0$
    \item the section is a parabola if and only if $\det A_{33} =0$
    \item the section is an ellipse if and only if $\det A_{33} >0$
\end{itemize}
If the conic section is degenerated ($\det A_Q = 0$), the conic section corresponds to:
\begin{itemize}
    \item 2 intersecting lines if and only if $\det A_{33} <0$
    \item 2 parallel straight lines if and only if $\det A_{33} =0$
    \item A single point if and only if $\det A_{33} >0$
\end{itemize}

Firstly, we look at the determinant of the $A_{Q1}$ matrix for the first equation of the system, corresponding to the first species:
\begin{equation}
    A_{Q1} = \begin{pmatrix}
    \alpha_{11}\beta_{11} & \frac{\alpha_{11}\beta_{12}+\alpha_{12}\beta_{11}}{2} & \frac{\beta_{11}+\alpha_{11}(1-C_1)}{2} \\
    \frac{\alpha_{11}\beta_{12}+\alpha_{12}\beta_{11}}{2} & \alpha_{12}\beta_{12} & \frac{\beta_{12}+\alpha_{12}(1-C_1)}{2} \\
    \frac{\beta_{11}+\alpha_{11}(1-C_1)}{2} & \frac{\beta_{12}+\alpha_{12}(1-C_1)}{2} & 1-C_1-D_1
    \end{pmatrix}.
\end{equation}

We can easily find an expression of the determinant of the matrix:
\begin{equation}
    \det A_{Q1} = (\alpha_{11}\beta_{12} - \alpha_{12}\beta_{11})D_1.
\end{equation}

We have $D_1 \neq 0$, therefore we have $\det A_{Q1} = 0 \Leftrightarrow \alpha_{11}\beta_{12} = \alpha_{12}\beta_{11}$.

We place ourselves in the case of $\alpha_{11}\beta_{12} = \alpha_{12}\beta_{11}$.
For the first equation of the system, we have :
\begin{equation}
    A_{33,1} = \begin{pmatrix}
    \alpha_{11}\beta_{11} & \frac{\alpha_{11}\beta_{12}+\alpha_{12}\beta_{11}}{2}\\
    \frac{\alpha_{11}\beta_{12}+\alpha_{12}\beta_{11}}{2} & \alpha_{12}\beta_{12}
    \end{pmatrix}.
\end{equation}
We can see that the determinant of the matrix $A_{33,1}$ is equal to 0 provided $\alpha_{11}\beta_{12} = \alpha_{12}\beta_{11}$. 
Then we find that the conic section corresponds to 2 parallel straight lines (a degenerate parabola) when $\alpha_{11}\beta_{12} = \alpha_{12}\beta_{11}$.

In the case where $\alpha_{11}\beta_{12} \neq \alpha_{12}\beta_{11}$, the conic section is non-degenerated, we have to look at the sign of $\det A_{33}$. For the first equation, we have:
\begin{align*}
    \det A_{33,1} &=  \alpha_{11}\beta_{11}\alpha_{12}\beta_{12} - (\frac{\alpha_{11}\beta_{12} + \alpha_{12}\beta_{11}}{2})^2\\
    &= \frac{1}{4}(\alpha_{11}\beta_{12}-\alpha_{12}\beta_{11})^2.
\end{align*}
The discriminant is then always positive, which indicates that the conic sections are hyperbola. As the system is symmetric, we have the same conclusions for species 2.\\

The analytical computation of the intersection of the conic sections, in the general case, requires a tedious analysis that consists in degenerating one conic section to find more easily the expression of the intersection.  Notably, it requires solving of a third degree equation that would lead to analytical results which are probably so complex that they might be impossible to interpret. 
We have not found a way to prove the uniqueness of a strictly positive intersection point but our graphical observations on multiple parameter sets seemed to show that we can find only one or zero positive point of intersection. 

In order to go further on the analysis of the intersection point, we placed ourselves in the case where $\alpha = \beta$. This case corresponds to a model where adult and juvenile parameters react in the same way to increases in adult densities. Even if it can be biologically inaccurate, it provides us a simple framework to understand how the conic sections can intersect.
With $\alpha=\beta$, we fulfill the condition to have degenerated conic sections, therefore the computation of the intersection points is easier. The equation of the degenerated conic section for species 1 becomes in this case:
\begin{equation}
    \alpha_{11}^2x^2+2\alpha_{11}\alpha_{12}xy + \alpha_{12}^2y^2 + \alpha_{11}(2-C_1)x + \alpha_{12}(2-C_1)y + (1-C_1-D_1) = 0.
\end{equation}
By separating the terms in $x$ and $y$, we can find the following expressions :
\begin{equation}
    y = -\frac{\alpha_{11}}{\alpha_{12}}x+\frac{\alpha_{11}}{\alpha_{12}}\left(\sqrt{\frac{1}{4}C_1^2+D_1}-1 + \frac{C_1}{2}\right)
\end{equation}
and
\begin{equation}
    y = -\frac{\alpha_{11}}{\alpha_{12}}x+\frac{\alpha_{11}}{\alpha_{12}}\left(-\sqrt{\frac{1}{4}C_1^2+D_1}-1 + \frac{C_1}{2}\right)
\end{equation}
which correspond to two parallel lines.
We have symmetrical equations for species 2, which allow us to compute the intersections between the two pairs of lines.
We find that for both species, only one of the two lines cross the section where $x>0$ and $y>0$. Therefore, there is at maximum one intersection point which has both coordinates positive. 
This point has for coordinates:
\begin{align}
    x &= \frac{\alpha_{12}\alpha_{21}}{\alpha_{22}\alpha_{12}-\alpha_{11}\alpha_{21}}\left(\frac{\alpha_{22}}{\alpha_{21}}\left(\sqrt{\frac{1}{4}C_2^2+D_2}-1 + \frac{C_2}{2}\right)-\frac{\alpha_{11}}{\alpha_{12}}\left(\sqrt{\frac{1}{4}C_1^2+D_1}-1 + \frac{C_1}{2}\right)\right)\\
    y &= \frac{\alpha_{11}\alpha_{22}}{\alpha_{22}\alpha_{12}-\alpha_{11}\alpha_{21}}\left(\sqrt{\frac{1}{4}C_1^2+D_1}+\frac{C_1}{2}-\sqrt{\frac{1}{4}C_2^2+D_2}-\frac{C_2}{2}\right).
\end{align}

We can search the conditions to have $x>0$ and $y>0$.\\
We assume $\alpha_{22}\alpha_{12}>\alpha_{11}\alpha_{21}$.\\
The condition to have $y>0$ is:
\begin{equation}
    \sqrt{\frac{1}{4}C_1^2+D_1}+\frac{C_1}{2}>\sqrt{\frac{1}{4}C_2^2+D_2}+\frac{C_2}{2}
\end{equation}
and to have $x>0$, we need:
\begin{align}
    \frac{\alpha_{22}}{\alpha_{21}}\left(\sqrt{\frac{1}{4}C_2^2+D_2}-1 + \frac{C_2}{2}\right)>\frac{\alpha_{11}}{\alpha_{12}}\left(\sqrt{\frac{1}{4}C_1^2+D_1}-1 + \frac{C_1}{2}\right)\\
    \alpha_{22}\alpha_{12}\left(\sqrt{\frac{1}{4}C_2^2+D_2}-1 + \frac{C_2}{2}\right) > \alpha_{11}\alpha_{21}\left(\sqrt{\frac{1}{4}C_1^2+D_1}-1 + \frac{C_1}{2}\right)
\end{align}
which is valid provided the condition to have $y>0$.

Therefore, we obtained the conditions to have a fixed point whose coordinates are both positive, which correspond to the conditions of coexistence, for the model where $\alpha = \beta$.\\

The search for fixed points in our model combining density-dependence on fertility and juvenile survival have allowed us to better understand how a strictly positive equilibrium (interior solution) can occur. With the simpler model where $\alpha=\beta$, we found additional conditions for coexistence than with the invasion analysis.
However, the analytical results remain difficult to interpret and/or restricted to the simplest models.

\newpage

\section{Invasion analysis for a model with metamorphosis life history structure}\label{SI:inv_analysis_MB}

We performed our invasion analysis on the model with a life-history structure similar to  \citet{moll2008competition} and \citet{cushing2007coexistence}, where juveniles and adults are separated, i.e. juveniles are in competition with juveniles and adults with adults. The projection matrix of this model is given in equation \eqref{eq:MB_model}.
The exact same method as in section \ref{sec:mat_met_inv_analysis} was used to compute the invasion criteria with this life-history structure.

In the metamorphosis model, with our notations and Beverton-Holt competition functions, the model can be written, for species 1:
\begin{equation}
    A_1^M = \begin{pmatrix}
    (1-\gamma_1)\frac{\phi_1}{1+\beta_{11}n_{1j}+\beta_{12}n_{2j}} & \frac{\pi_1}{1+\alpha_{11}n_{1a}+\alpha_{12}n_{2a}} \\
    \gamma_1\frac{\phi_1}{1+\beta_{11}n_{1j}+\beta_{12}n_{2j}} & s_{1a}
    \end{pmatrix}.
\end{equation}

We applied the invasion analysis as for our combined model. Firstly, we search the densities of exclusion equilibria. Then, we have the following system for the model with a single species (we drop here the time indices for clarity and because the model is evaluated at the fixed point):
\begin{equation}\label{eq:system_single_sp_MB}
    \left\{
\begin{array}{r c l}
n_j &=& (1-\gamma)\frac{\phi}{1+\beta n_j}n_j + \frac{\pi}{1+\alpha n_a}n_a\\
n_a &=& \gamma \frac{\phi}{1+\beta n_j}n_j + s_a n_a.
\end{array}
\right.
\end{equation}

We can easily express $n_j$ according to $n_a$:
\begin{align*}
    n_a = \frac{\gamma\phi}{1+\beta n_j}n_j + s_a n_a\\
    \Leftrightarrow \gamma\phi n_j = n_a(1-s_a)+n_a(1-s_a)\beta n_j\\
    \Leftrightarrow (\gamma\phi - n_a (1-s_a)\beta)n_j = n_a(1-s_a)\\
\end{align*}
so that 
\begin{equation}
    n_j = \frac{(1-s_a)n_a}{\gamma\phi - (1-s_a)\beta n_a}.
\end{equation}

We replace $n_j$ by its expression in the first equation of the system \eqref{eq:system_single_sp_MB}, that leads to the polynomial of degree 2:
\begin{equation}
    E\alpha\beta n_a^2 + \left(\alpha(1-C) +\beta(\pi+E)\right)n_a +1-C-D
\end{equation}
where $C = \phi(1-\gamma)$, $D=\frac{\pi\gamma\phi}{1-s_a}$ and $E=\frac{(1-\gamma)(1-s_a)}{\gamma}$.\\

We find the same condition for inherent net reproductive number than for our model 1. It is given by $C+D >1$ that must be true to have a viable species. This is due to the similarity of the equations to those of model 1, which mainly differ in that factor $E$ now appears in front of the degree 2 term. We then have positive densities for the exclusion equilibrium where one species is absent:
\begin{equation}
    \left\{
\begin{array}{r c l}
n_j^* &=&  \frac{(1-s_a)n_a^*}{\gamma\phi - (1-s_a)\beta n_a^*}\\
n_a^* &=& \frac{-\left(\alpha(1-C) +\beta(\pi+E)\right)+\sqrt{\left(\alpha(1-C) +\beta(\pi+E)\right)^2-4E\alpha\beta(1-C-D)}}{2E\alpha\beta}
\end{array}
\right.
\end{equation}
provided 
\begin{equation}
   C+D = (1-\gamma)\phi +\frac{\pi \gamma \phi}{1-s_a} > 1. 
\end{equation}

We have then obtained the densities at the exclusion equilibrium. We have now to evaluate the stability of this equilibrium: if it is locally asymptotically stable, the excluded species cannot invade the community starting from very small density. Conversely, if it is unstable, this means that the excluded species can invade \textit{provided that equilibria of the single-species model are restricted to stable fixed points}. This last restriction is key. Indeed, if the single-species model has a stable two-cycle or an even more complex fluctuating attractor, the fixed point will already be unstable, and a proper invasion analysis should instead be done on this more complex attractor. Two-cycles tend to occur for $s_a=0$ \citep{cushing2007coexistence}, as shown in \ref{SI:simuls_twospecies}. So long as $s_a>0$ this model does not show 2-cycles---so far as we have seen numerically and following the argumentation of \citet{cushing2007coexistence} for model 2. We therefore restrict ourselves to the case $s_a>0$, although the equations derived above in this Appendix for the single-species fixed point are also valid whenever $s_a=0$.

To evaluate the (un)stability of the exclusion equilibria in the two-species system, we have to compute the eigenvalues of the Jacobian of the 2 species system evaluated at the exclusion equilibrium. The full system iterated over a time step reads:
\begin{equation}
    \left\{
\begin{array}{r c l}
n_{1j}(t+1) &=& (1-\gamma_1)\frac{\phi_1}{1+\beta_{11}n_{1j}(t)+\beta_{12}n_{2j}(t)}n_{1j}(t) + \frac{\pi_1}{1+\alpha_{11}n_{1a}(t)+\alpha_{12}n_{2a}(t)}n_{1a}(t)\\
n_{1a}(t+1) &=& \gamma_1 \frac{\phi_1}{1+\beta_{11}n_{1j}(t)+\beta_{12}n_{2j}(t)}n_{1j}(t) + s_{1a}n_{1a}(t)\\
n_{2j}(t+1) &=& (1-\gamma_2)\frac{\phi_2}{1+\beta_{22}n_{2j}(t)+\beta_{21}n_{1j}(t)}n_{2j}(t) + \frac{\pi_2}{1+\alpha_{22}n_{2a}(t)+\alpha_{21}n_{1a}(t)}n_{2a}(t)\\
n_{2a}(t+1) &=& \gamma_2 \frac{\phi_2}{1+\beta_{22}n_{2j}(t)+\beta_{21}n_{1j}(t)}n_{2j}(t) + s_{2a}n_{2a}(t).
\end{array}
\right.
\end{equation}

We place ourselves at the abovementioned exclusion equilibrium where species 1 dominates the community and species 2 is absent. The Jacobian evaluated at this point is a $4 \times 4$ matrix and it has the following triangular block form:
\begin{equation}
    \begin{pmatrix}
    B_1 & B_2 \\
    0 & B_3 \\
    \end{pmatrix}.
\end{equation}
Therefore, we only need to know the eigenvalues of the $2 \times 2$ matrices $B_1$ and $B_3$. 
The $B_1$ matrix corresponds to the Jacobian of the single-species dynamical system, which has always been found to be stable numerically, thus for reasons explained in more detail in \ref{app:inv_analysis} we assume the eigenvalues have a modulus smaller than 1.
The $B_3$ matrix is given by:
\begin{equation}
    B_3 =
    \begin{pmatrix}
    (1-\gamma_2)\frac{\phi_2}{1+\beta_{21}n_{1j}^*} & \frac{\pi_2}{1+\alpha_{21}n_{1a}^*} \\
    \gamma_2\frac{\phi_2}{1+\beta_{21}n_{1j}^*} & s_{2a} \\
    \end{pmatrix}.
\end{equation}
The $B_3$ matrix corresponds to the projection matrix of species 2, which implies that the exclusion equilibrium is not stable when species 2 has a growth rate larger than 1 (i.e., the modulus of $B_3$'s leading eigenvalue is larger than 1). The stability of this two-dimensional system can be investigated thanks to the Jury conditions, given by the equations (\ref{1st_jury_condition}, \ref{2nd_jury_condition}, \ref{3rd_jury_condition}):

\begin{align}
1- \tr J(\mathbf{n})+\det J(\mathbf{n}) &>0\label{1st_jury_condition}\\ 
1+ \tr J(\mathbf{n}) + \det J(\mathbf{n}) &> 0 \label{2nd_jury_condition}\\
1- \det J(\mathbf{n}) &> 0 \label{3rd_jury_condition}
\end{align}

with $J(\mathbf{n})$ the Jacobian matrix of the two-compartment system whose stability is being evaluated. If these three conditions hold, the system is stable. If the first condition (equation \eqref{1st_jury_condition}) is violated, one of the eigenvalues of $J(\mathbf{n})$ is larger than 1, which means in our context that the excluded species can invade the community. The violations of the conditions \eqref{2nd_jury_condition} and \eqref{3rd_jury_condition} correspond to the creation of limit cycles \citep{neubert2000density}. We then check these conditions on the $B_3$ matrix. We have:
\begin{align*}
    \tr B_3 &= (1-\gamma_2)\frac{\phi_2}{1+\beta_{21}n_{1j}^*} + s_{2a} \\
    \det B_3 &= (1-\gamma_2)\frac{\phi_2}{1+\beta_{21}n_{1j}^*}s_{2a} - \gamma_2\frac{\phi_2}{1+\beta_{21}n_{1j}^*}\frac{\pi_2}{1+\alpha_{21}n_{1a}^*}.
\end{align*}
We start with the third condition (\ref{3rd_jury_condition}):
\begin{equation}\label{eq:Jury3}
    1- (1-\gamma_2)\frac{\phi_2}{1+\beta_{21}n_{1j}^*}s_{2a} + \gamma_2\frac{\phi_2}{1+\beta_{21}n_{1j}^*}\frac{\pi_2}{1+\alpha_{21}n_{1a}^*} > 0.
\end{equation}
This equation will always be satisfied with biologically meaningful parameters and strictly competitive interaction between species: 
\begin{itemize}
    \item The third term of \eqref{eq:Jury3} is positive because each parameter is positive,
    \item $(1-\gamma_2)\frac{\phi_2}{1+\beta_{21}n_{1j}^*}s_{2a} < 1$ because $0 \leq \gamma \leq 1$, $0 \leq \phi \leq 1$ and all $\beta$'s are positive.
\end{itemize}

Since $\tr B_3 > 0$, the first Jury's condition implies the second ((\ref{1st_jury_condition}) $\Rightarrow$ (\ref{2nd_jury_condition})). Then, we only have to check the first Jury condition to describe the stability of the system:

\begin{align*}
    1-(1-\gamma_2)\frac{\phi_2}{1+\beta_{21}n_{1j}^*} + s_{2a}+(1-\gamma_2)\frac{\phi_2}{1+\beta_{21}n_{1j}^*}s_{2a} - \gamma_2\frac{\phi_2}{1+\beta_{21}n_{1j}^*}\frac{\pi_2}{1+\alpha_{21}n_{1a}^*} > 0 \\
    (1-\gamma_2)\frac{\phi_2}{1+\beta_{21}n_{1j}^*} - (1-\gamma_2)\frac{\phi_2}{1+\beta_{21}n_{1j}^*}s_{2a} + \gamma_2\frac{\phi_2}{1+\beta_{21}n_{1j}^*}\frac{\pi_2}{1+\alpha_{21}n_{1a}^*} < 1-s_{2a}\\
    (1-\gamma_2)\phi_2(1-s_{2a}) < (1-s_{2a})(1+\beta_{21}n_{1j}^*)- \frac{\gamma_2\phi_2\pi_2}{1+\alpha_{21}n_{1a}^*}\\
    (1-\gamma_2)\phi_2\frac{1}{1+\beta_{21}n_{1j}^*} + \frac{\gamma_2\phi_2\pi_2}{1-s_{2a}}\frac{1}{(1+\beta_{21}n_{1j}^*)(1+\alpha_{21}n_{1a}^*)} < 1.
\end{align*}

We finally have the following condition for the stability of the equilibrium:
\begin{equation}
    \frac{C_2}{1+\beta_{21}n_{1j}^*}+\frac{D_2}{(1+\beta_{21}n_{1j}^*)(1+\alpha_{21}n_{1a}^*)} < 1
\end{equation}
with $C_2= (1-\gamma_2)\phi_2$ and $D_2= \frac{\pi_2\gamma_2\phi_2}{1-s_{2a}}$.\\

\begin{figure}[H]
    \centering
    \includegraphics[width=0.8\textwidth]{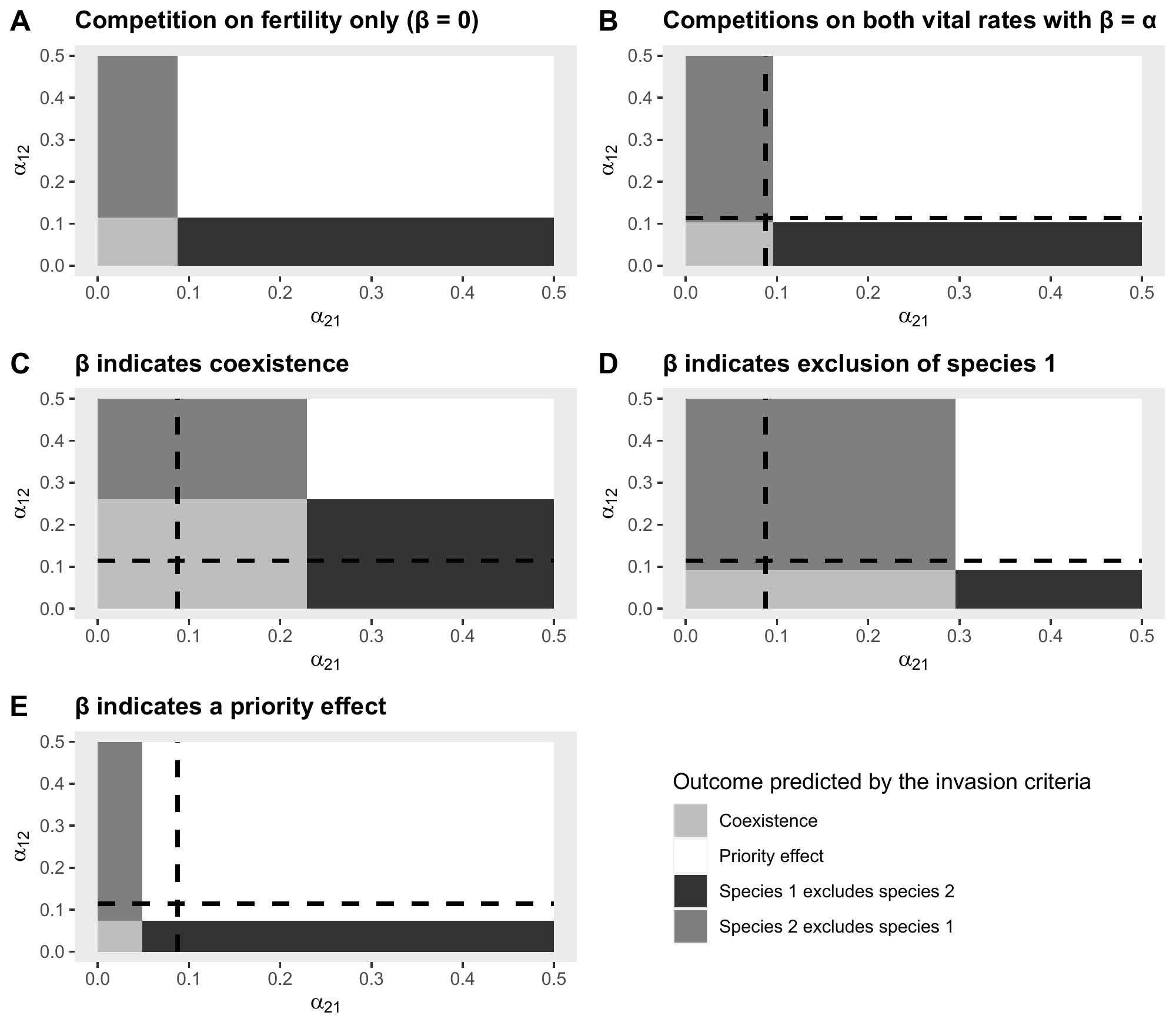}
    \caption{Outcomes of the model with a metamorphosis structure predicted by invasion analysis, depending on the coefficients of inter-specific competition on fertility ($\alpha_{12}$ and $\alpha_{21})$ with $\beta$ coefficients indicating coexistence, exclusion of species 1 and priority effect. The dotted lines correspond to the outcome boundaries of panel A which is used as a reference.}
    \label{fig:MB_invasion_outcome}
\end{figure}

We obtain the coexistence outcome domains given in Fig. \ref{fig:MB_invasion_outcome} with the $\beta$ coefficients given in Table \ref{tab:beta_inv_analysis_MB}. The results are qualitatively similar to those presented in the main text with adults $\rightarrow$ adults + juveniles.

\begin{table}[H]
    \centering
\begin{tabular}{|c|rr|rr|}
  \hline
Figure & $\beta_{12}$ & $\beta_{21}$ \\ 
  \hline
C & 0.06 & 0.06 \\ 
D & 0.11 & 0.05\\ 
E & 0.12 & 0.12\\ 
   \hline
\end{tabular}
\caption{$\beta$ competition coefficients used to obtain contrasting situations of coexistence.}
\label{tab:beta_inv_analysis_MB}
\end{table}

\newpage

\section{Dynamics of single-species and two-species models}\label{SI:simuls_twospecies}
Here we explore the dynamics of single- and two-species models in four contrasted life-history configurations (identical to those previously considered by \citealt{neubert2000density} in single-species models where one vital rate has Ricker density-dependence):
\begin{itemize}
    \item $s_a>0, \gamma<1$ which corresponds to the parameters used for illustration in the main text
    \item $s_a>0, \gamma=1$: iteroparity with fast maturation
    \item $s_a=0, \gamma<1$: semelparity and delayed maturation
    \item $s_a=0, \gamma=1$: semelparity and fast maturation.
\end{itemize}
For single-species models (which we use when performing invasion analysis at the exclusion equilibria), only the last case (i.e. semelparity and fast maturation) leads to 2-cycles with the remainder of parameters as in the main text (see figures below). This stays true when multiplying the fecundity by 10---a choice motivated by the fact that high fecundities generally enhance cycling. For the metamorphosis model, the semelparity and fast maturation case leads to a fixed point when initial conditions are strictly positive for juveniles and adults. Indeed, \citet{cushing2007coexistence} covers the semelparous, fast maturation version of metamorphosis model and showed that while the nontrivial equilibrium for the single-species model is stable on the positive cone, it is not so when starting on the boundary with either $n_j=0$ or $n_a=0$, which can generate so-called synchronous two-cycles (Theorem 2.1, \citealt{cushing2007coexistence}).
 
For two-species models, the case $s_a=0, \gamma=1$ generates 2-cycles but $s_a=0, \gamma<1$ can also generate 2-cycles for some parameter sets, and some different kinds of cycles (see below). \\ 

\begin{figure}[H]
    \centering
    \includegraphics[width=0.8\textwidth]{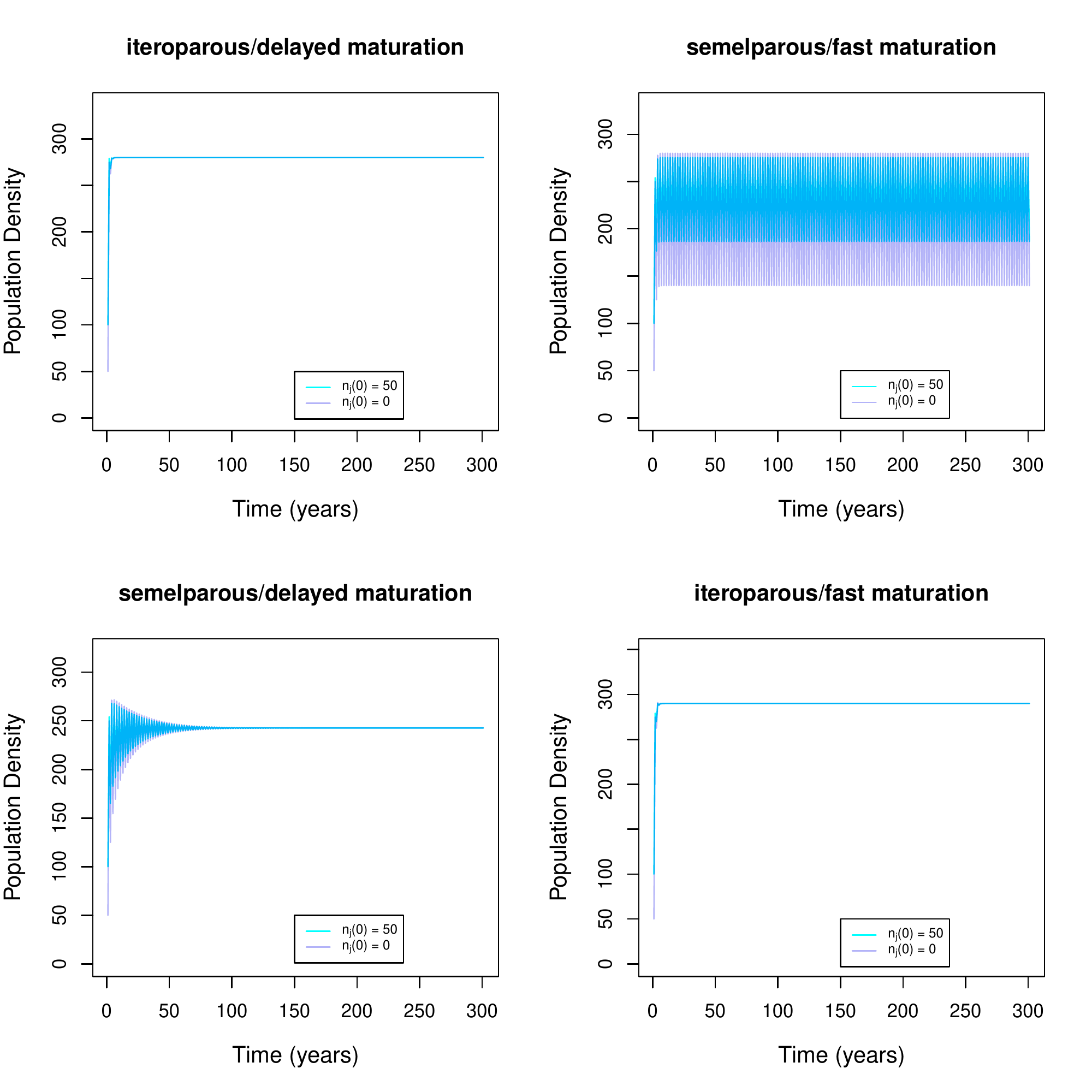}
    \caption{Population dynamics for the single species model and competition generated by adults (model 1) with $n_{j}(0) = 50$ and $n_{a}(0) = 50$ (in cyan) and for $n_{j}(0) = 0$ and $n_{a}(0) = 50$ (in purple). Similar dynamics are obtained with both initial conditions for model 1. }
    \label{fig:1sp_dynamics}
\end{figure}

\begin{figure}[H]
    \centering
    \includegraphics[width=0.8\textwidth]{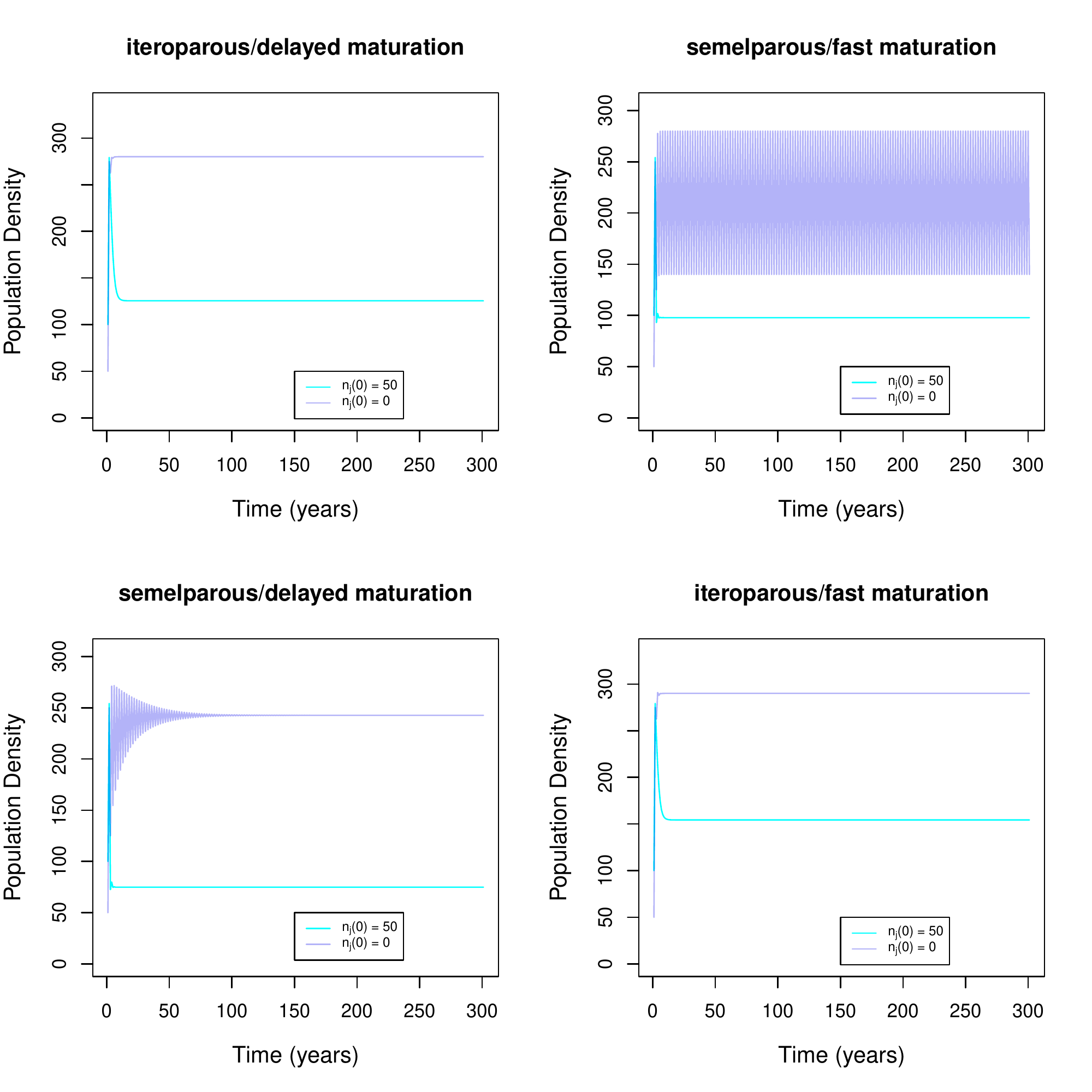}
    \caption{Population dynamics for the single species model and metamorphosis structure with $n_{j}(0) = 50$ and $n_{a}(0) = 50$ (in cyan) and for $n_{j}(0) = 0$ and $n_{a}(0) = 50$ (in purple).  A stable fixed point is always obtained with $n(0)=(50,50)$ for model 2. }
    \label{fig:1sp_dynamics_meta}
\end{figure}

\begin{figure}[H]
    \centering
    \includegraphics[width=0.8\textwidth]{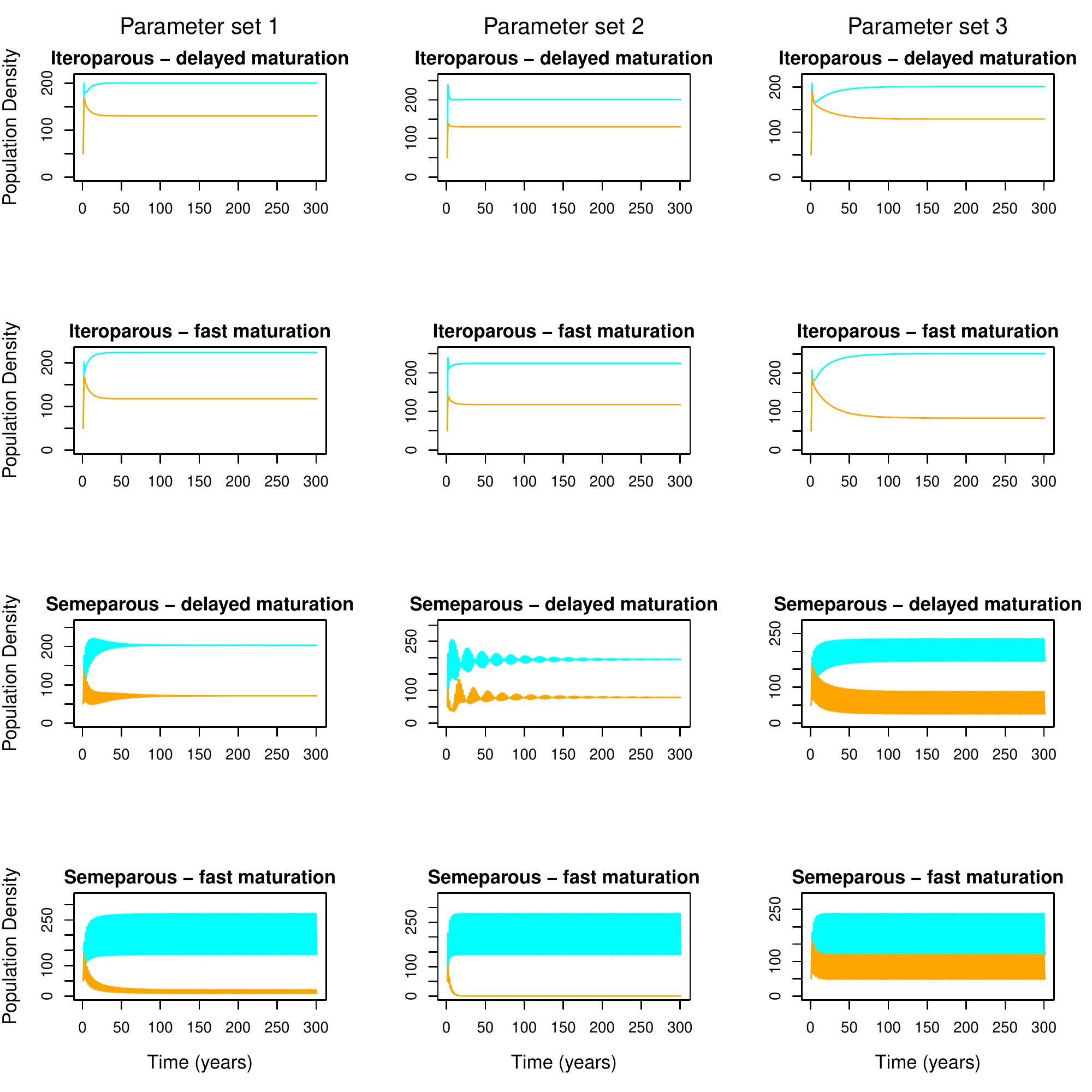}
    \caption{Population dynamics for the two-species models 1, where competition is generated by adults,  with the three parameter sets built to generate three different scenarios of coexistence. We consider 4 different life history configurations, always with the initial conditions $n_i(0) = (0,50)$ to allow for all accessible 2-cycles.}
    \label{fig:2sp_dynamics}
\end{figure}
For the reader interested in dynamical aspects, we recommend to consult \citet{cushing2007coexistence}. To facilitate the reading, we make a correspondence below between their parameter notations and ours. Their model writes
\begin{equation}
\begin{split}
       J_{t+1} = b_1 \frac{1}{1+d_1A_t} A_t\\
       A_{t+1} = s_1 \frac{1}{1+ J_t + c_1 j_t} J_t\\
       j_{t+1} = b_2 \frac{1}{1+d_2 a_t} a_t\\
       a_{t+1} = s_2 \frac{1}{1+c_2 J_t + j_t} j_t
\end{split}
\end{equation}
with a state-space vector $(J_t,A_t, j_t,a_t)$ where one of the species is written in capitals. \\

\citet{cushing2007coexistence} is therefore a special case of 
\begin{equation}
    \mathbf{A}_1^{M}(\mathbf{n}) = \begin{pmatrix}
    \frac{(1-\gamma_1)\phi_1}{1+\beta_{11}n_{1j}+\beta_{12}n_{2j}} & \frac{\pi_1}{1+\alpha_{11}n_{1a}+\alpha_{12}n_{2a}} \\
    \frac{\gamma_1\phi_1}{1+\beta_{11}n_{1j}+\beta_{12}n_{2j}} & s_{1a}
    \end{pmatrix}\label{eq:MB_model}
\end{equation}
that is,
\begin{equation}
    \mathbf{A}_1^{M}(\mathbf{n}) = \begin{pmatrix}
    0 & \frac{\pi_1}{1+\alpha_{11}n_{1a}+0 \times n_{2a}} \\
    \frac{\phi_1}{1+\beta_{11}n_{1j}+\beta_{12}n_{2j}} & 0
    \end{pmatrix}.\label{eq:MB_model}
\end{equation}
Note that $\alpha_{11} = d_1, \alpha_{12} = \alpha_{21} = 0, \alpha_{22} = d_2, \pi_1 = b_1, \pi_2 = b_2, \beta_{11} = 1, \beta_{12} = c_1, \beta_{21} = c_2, \beta_{22} = 1, \phi_1 = s_1,\phi_2 = s_2$.

\subsection*{Why is this sub-case of the ``metamorphosis life cycle'' model more amenable to analytical insights?}

\paragraph{Going to $t+2$}

\begin{equation}
    n_{1j}(t+2) =  \frac{\pi_1 {\color{blue}n_{1a}(t+1)}}{1+\alpha_{11}{\color{blue}n_{1a}(t+1)}+0 \times n_{2a}(t+1)}
\end{equation}
with $\color{blue}n_{1a}(t+1) = \frac{\phi_1 n_{1j}(t)}{1+\beta_{11}n_{1j}(t)+\beta_{12}n_{2j}(t)}$. This leads to 

\begin{equation}
    n_{1j}(t+2) = \frac{\pi_1 {\color{blue} \frac{\phi_1 n_{1j}(t)}{1+\beta_{11}n_{1j}(t)+\beta_{12}n_{2j}(t)}}}{1+\alpha_{11}{\color{blue} \frac{\phi_1 n_{1j}(t)}{1+\beta_{11}n_{1j}(t)+\beta_{12}n_{2j}(t)}}}
\end{equation}
which, multiplying by $1+\beta_{11}n_{1j}(t)+\beta_{12}n_{2j}(t)$ the numerator and denominator, leads to 
\begin{equation}
\begin{split}
        n_{1j}(t+2) = \frac{\pi_1 \phi_1 n_{1j}(t)}{1+\beta_{11}n_{1j}(t)+\beta_{12}n_{2j}(t) +\alpha_{11}\phi_1 n_{1j}(t)}\\
        = \frac{\pi_1 \phi_1 n_{1j}(t)}{1+(\beta_{11}+\alpha_{11}\phi_1) n_{1j}(t)+\beta_{12}n_{2j}(t)}. 
\end{split}
\end{equation}

This equation gives $n_{1j}(t+2)$ as a function of $n_{1j}(t)$ and $n_{2j}(t)$ and is equivalent to the classical 2-species Leslie-Gower model (except for the doubled timestep), and therefore we know its behaviour (fixed point), as shown by \citet{cushing2007coexistence}.  Because $n_{1a}(t+1)$ only depends on $n_{1j}(t)$ and $n_{2j}(t)$ we can deduce that it won't change the dynamical behaviour.

\newpage

\section{Simulations of metamorphosis model structure (model 2) extended to a $S$-species context}\label{SI:sim_model_MB}
\begin{figure}[H]
    \centering
    \includegraphics[scale=0.9]{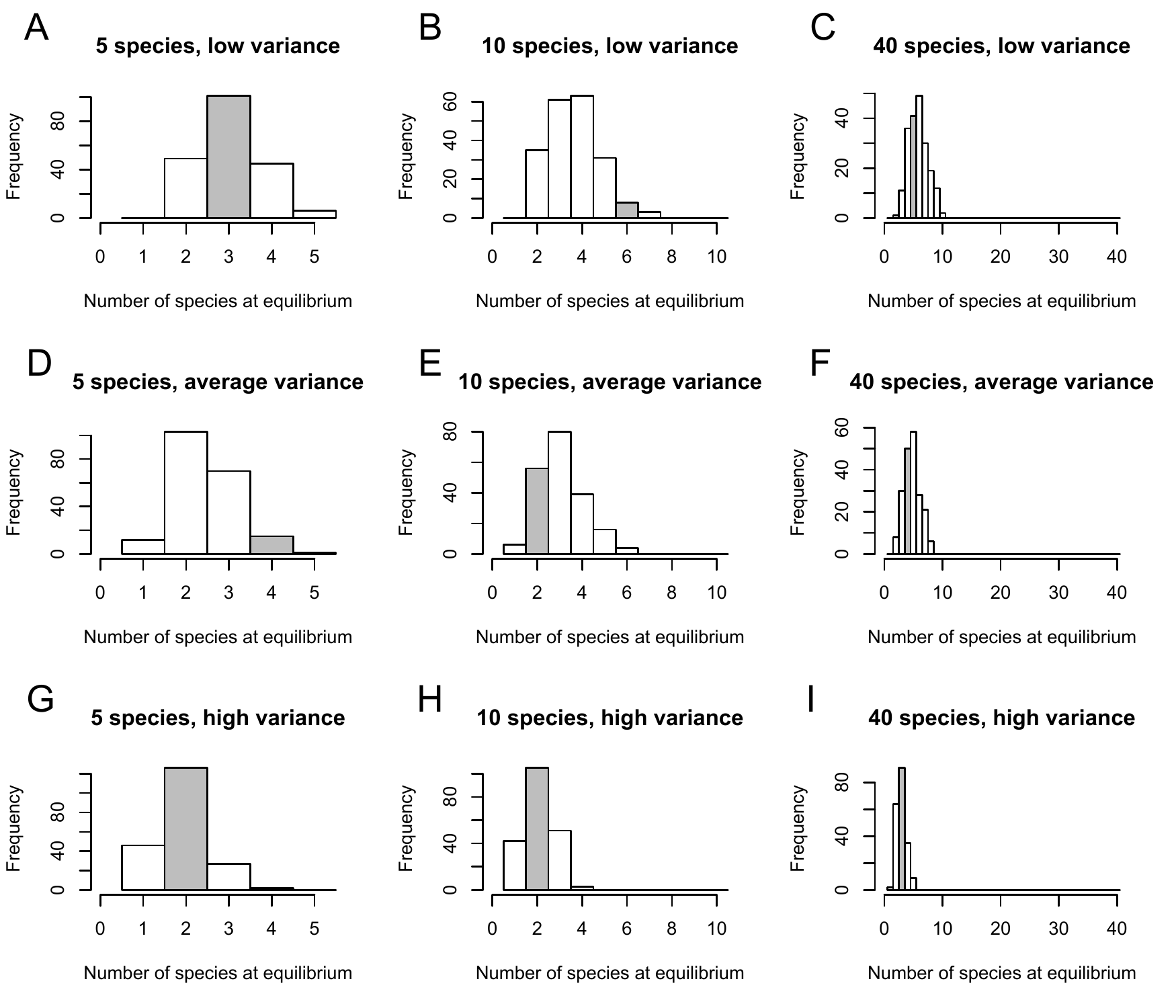}
    \caption{Histograms of the number of species persisting at equilibrium for 200 permutations of inter-specific competition coefficients. The bar in gray indicates the number of species persisting at time $t=3000$ time steps for the original $\alpha_{ij}$ and $\beta_{ij}$ with $\text{Corr}(\alpha_{ij}, \beta_{ij})<0$. From left to right, columns correspond the communities with initially 5, 10 and 40 species. From top to bottom, the lines correspond to situations with high, average and low variance on the vital rates.}
    \label{fig:hist_opp_hier_only_MB}
\end{figure}

The species richness at the end of simulations for the extended model 2 with $S$ species are given in Fig. \ref{fig:hist_opp_hier_only_MB}. 

We find here that with a slightly different competition structure (adults only compete with adults and juveniles with juveniles), results are only slightly different from those of model 1. They go in the same general direction: it is difficult to obtain a positive effect of opposite competitive hierarchies on species richness at equilibrium. One could have believed that there was something special with \citet{moll2008competition}'s life cycle structure with separated adults from juveniles, which would have made many-species coexistence despite competition a little bit easier. We found the opposite: for this particular set of simulations, we have not been able to observe more coexistence with opposite competitive hierarchies on $\alpha$ and $\beta$, even for very similar species. This only strengthens the generality of our finding that opposite competitive hierarchies are unlikely to be a general explanation to many-species coexistence. \end{document}